\documentclass[aip,rsi,reprint,floatfix]{revtex4-1}

\usepackage{graphicx}
\usepackage{upgreek}
\usepackage[pdftex,bookmarks,colorlinks,allcolors=blue]{hyperref}
\usepackage{natbib}
\usepackage[english]{babel}
\usepackage{ifthen}
\usepackage{booktabs}
\usepackage{nomencl}
\newcommand{\f}[1]{\ensuremath{\mathrm{f}_{#1}}} 		
\newcommand{\fexp}{\ensuremath{{f_\mathrm{exp}}}} 		
\newcommand{\Do}{\ensuremath{d_\mathrm{L}}} 						
\newcommand{\Ro}{\ensuremath{R_0}} 						
\newcommand{\tauc}{\ensuremath{\tau_c}} 				
\newcommand{\taup}{\ensuremath{\tau_p}} 				
\newcommand{\We}{\ensuremath{\mathrm{We}}}				
\newcommand{\Rey}{\ensuremath{\mathrm{Re}}}				
\newcommand{\ELaser}{\ensuremath{E_\mathrm{L}}}			
\newcommand{\lLaser}{\ensuremath{\lambda_\mathrm{L}}}	
\newcommand{\Fig}[1][]{%
\ifthenelse{\equal{#1}{}}{Fig.~}{Fig#1.~}%
}
\newcommand{\Chap}[1][]{%
\ifthenelse{\equal{#1}{}}{\S\,}{\S#1\,}%
}
\newcommand{\Eq}[1][]{%
\ifthenelse{\equal{#1}{}}{Eq.~}{Eq#1.~}%
}
\newcommand{\Tab}[1][]{%
\ifthenelse{\equal{#1}{}}{Table~}{Table#1~}%
}

\makeatletter
\def\bbl@set@language#1{%
  \edef\languagename{%
    \ifnum\escapechar=\expandafter`\string#1\@empty
    \else\string#1\@empty\fi}%
  \@ifundefined{babel@language@alias@\languagename}{}{%
    \edef\languagename{\@nameuse{babel@language@alias@\languagename}}%
  }%
  \select@language{\languagename}%
  \expandafter\ifx\csname date\languagename\endcsname\relax\else
    \if@filesw
      \protected@write\@auxout{}{\string\select@language{\languagename}}%
      \bbl@for\bbl@tempa\BabelContentsFiles{%
        \addtocontents{\bbl@tempa}{\xstring\select@language{\languagename}}}%
      \bbl@usehooks{write}{}%
    \fi
  \fi}
\newcommand{\DeclareLanguageAlias}[2]{%
  \global\@namedef{babel@language@alias@#1}{#2}%
}
\makeatother

\DeclareLanguageAlias{en}{english}

\makeatletter
\let\Hy@backout\@gobble
\makeatother
\newcommand{\refl}[1]{\protect\includegraphics{#1}}

\AtBeginDocument{
\heavyrulewidth=.08em
\lightrulewidth=.05em
\cmidrulewidth=.03em
\belowrulesep=.65ex
\belowbottomsep=0pt
\aboverulesep=.4ex
\abovetopsep=0pt
\cmidrulesep=\doublerulesep
\cmidrulekern=.5em
\defaultaddspace=.5em
}

\begin{document} 
\title{Apparatus to control and visualize the impact of a high-energy laser pulse on a liquid target} 
\author{Alexander L. Klein} \email[]{alexludwigklein@gmail.com}
\affiliation{Physics of Fluids Group, Max Planck Center Twente for Complex Fluid Dynamics, JM Burgers Center, and MESA+ Center for Nanotechnology, Department of Science and Technology, University of Twente, P.O. Box 217, 7500 AE Enschede, The Netherlands.}
\author{Detlef Lohse} 
\affiliation{Physics of Fluids Group, Max Planck Center Twente for Complex Fluid Dynamics, JM Burgers Center, and MESA+ Center for Nanotechnology, Department of Science and Technology, University of Twente, P.O. Box 217, 7500 AE Enschede, The Netherlands.}
\affiliation{Max Planck Institute for Dynamics and Self-Organization, Am Fassberg 17, 37077 G\"ottingen, Germany.}
\author{Michel Versluis} 
\affiliation{Physics of Fluids Group, Max Planck Center Twente for Complex Fluid Dynamics, JM Burgers Center, and MESA+ Center for Nanotechnology, Department of Science and Technology, University of Twente, P.O. Box 217, 7500 AE Enschede, The Netherlands.}
\author{Hanneke Gelderblom} 
\affiliation{Physics of Fluids Group, Max Planck Center Twente for Complex Fluid Dynamics, JM Burgers Center, and MESA+ Center for Nanotechnology, Department of Science and Technology, University of Twente, P.O. Box 217, 7500 AE Enschede, The Netherlands.}
\date{\today}

\begin{abstract}
We present an experimental apparatus to control and visualize the response of a liquid target to a laser-induced vaporization. We use a millimeter-sized drop as target and present two liquid-dye solutions that allow a variation of the absorption coefficient of the laser light in the drop by seven orders of magnitude. The excitation source is a Q-switched Nd:YAG laser at its frequency-doubled wavelength emitting nanosecond pulses with energy densities above the local vaporization threshold. The absorption of the laser energy leads to a large-scale liquid motion at timescales that are separated by several orders of magnitude, which we spatiotemporally resolve by a combination of ultra-high-speed and stroboscopic high-resolution imaging in two orthogonal views. Surprisingly, the large-scale liquid motion at upon laser impact is completely controlled by the spatial energy distribution obtained by a precise beam-shaping technique. The apparatus demonstrates the potential for accurate and quantitative studies of laser-matter interactions.
\end{abstract}

\maketitle

\makenomenclature
{\small
\printnomenclature[0.6in]
}
\nomenclature{AOI}{angle of incident}
\nomenclature{BP}{beam profile}
\nomenclature{CASRN}{chemical abstracts service registry number}
\nomenclature{CCD}{charge-coupled device}
\nomenclature{EM}{energy meter}
\nomenclature{EUV}{extreme ultraviolet}
\nomenclature{FL}{flash lamp}
\nomenclature{FOV }{field of view}
\nomenclature{fps}{frames per second}
\nomenclature{FWHM}{full width at half maximum}
\nomenclature{iLIF}{incoherent laser-induced fluorescence}
\nomenclature{LDM}{long-distance microscope}
\nomenclature{LIDT}{laser-induced damage threshold}
\nomenclature{MEK}{methyl ethyl ketone}
\nomenclature{ND}{neutral density}
\nomenclature{Nd:YAG}{neodymium-doped yttrium aluminum garnet}
\nomenclature{OD}{optical density}
\nomenclature{PBS}{polarizing beam splitter}
\nomenclature{PD}{photodiode}
\nomenclature{QS}{Q-switch}
\nomenclature{SHG}{second harmonic generator}
\nomenclature{SLR}{single-lens reflex}
\nomenclature{TL}{trigger laser}
\nomenclature{TTL}{transistor-transistor logic}
\nomenclature{USB}{universal serial bus}

\section{Introduction}\label{sec:Introduction}
\begin{figure*}[t!]
	\begin{minipage}[t]{0.50\textwidth}
		\mbox{}\\
		\includegraphics[width=10.5cm]{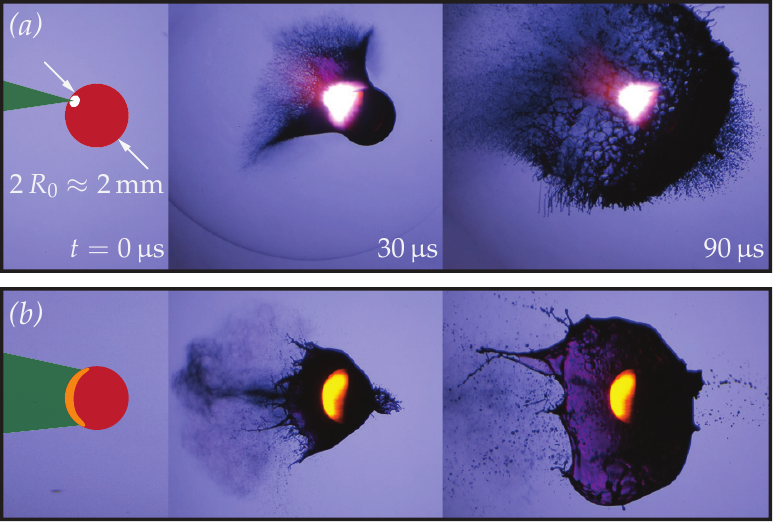}
		\end{minipage}\hfill
		\begin{minipage}[t]{0.39\textwidth}
			\mbox{}\\[-\baselineskip]
			\caption{Laser-induced liquid motion in side-view images taken stroboscopically with a color camera and a broadband pulsed light source for two laser focusing conditions (illustrated by a sketch in the left column): (a) Tightly focused laser beam resulting in plasma formation (visible as white glow) and a violent ablation from the drop. A spherical shockwave is visible at $t=30\,\upmu\mathrm{s}$. (b) Moderately focused laser beam leading to vaporization in absence of any plasma (fluorescence emission of the dye molecules is visible as yellow spot). The liquid used in the experiments is an aqueous magenta-colored ink that absorbs the laser light emitted at a wavelength of $\lLaser=532\,\mathrm{nm}$. In both cases, the recoil pressure of the phase change imparts momentum to the drop. As a response, the initially spherical drop evolves into a strongly curved and thin liquid sheet that breaks into tiny fragments.\label{fig:ImpactRegimes}}
		\end{minipage}
	\end{figure*}

Light moves liquid matter in various ways. First of all, the direct interaction of an electromagnetic wave with a liquid surface exerts a pressure on the interface that may move the liquid: photons exchange momentum with the liquid as the light path changes at the interface between two media of different refractive index, an effect described as optical radiation pressure~\cite{ashkin_radiation_1973}.
The observed motion is usually small~\cite{sakai_measurement_2001} unless competing forces such as capillary forces are weakened, for example under near-critical conditions when surface tension vanishes~\cite{casner_giant_2001}. 
Intense field strengths are required to exert the radiation pressure onto the liquid target, and therefore focused laser light is used.
Second, laser radiation can also induce thermocapillary forces, 
which can be applied to control liquids. The localized heating by a laser, either directly by linear absorption of the light in the liquid sample or indirectly by heating the substrate with which the liquid sample is in contact, introduces thermocapillary stresses~\cite{viznyuk_thermocapillary_1988, chraibi_thermocapillary_2012}. 
This effect of induced liquid motion by thermocapillary stresses is well-known as the Marangoni effect~\cite{marangoni_uber_1871}. A laser-induced phase change is a third way to move liquids by optical radiation, which allows for large deformations and flow speeds to be reached, see~\Fig\ref{fig:ImpactRegimes}. The liquid motion driven by this method is the phenomena under investigation in the experimental apparatus that we present in this paper.

A laser-induced phase change requires the local field strength or energy density to be high enough to supply the energy required for a phase change such as vaporization or plasma formation.
Local vaporization is achieved by linear absorption of the laser energy in the liquid, or by additives, e.g.\ dyes or dyed particles~\cite{kafalas_fog_1973, kafalas_dynamics_1973, pinnick_micron-sized_1990, apitz_material_2005, sun_growth_2009, tagawa_highly_2012, lajoinie_ultrafast_2014}.
Self-focusing and dielectric breakdown may lead to plasma formation in otherwise transparent liquids~\cite{zhang_explosive_1987, vogel_shock_1996, favre_white-light_2002, lindinger_time-resolved_2004, vogel_mechanisms_2005, thoroddsen_spray_2009, geints_broadband_2010}.
Both effects are for example exploited for medical applications in biological matter~\cite{vogel_mechanisms_2003,gattass_femtosecond_2008}.\@ An application where the target material is initially solid and needs to melt, i.e.\ the liquid phase is an intermediate step, is found in the laser-induced forward transfer of metals~\cite{visser_toward_2015}.
The preparation of liquid and solid samples for mass spectroscopy by laser-induced desorption or ionization~\cite{merchant_recent_2000,norris_analysis_2013} are examples where the phase change itself is of primary interest and the induced motion is a secondary effect. Similarly, in pulsed laser deposition~\cite{eason_pulsed_2007}, vaporization is the primary method to transfer the material of interest, including liquids~\cite{rong_liquid_1995}, to a thin film on a substrate.

The setup presented here is motivated by the physical processes and especially the fluid dynamics found in laser-produced plasma light sources for extreme ultraviolet (EUV) nanolithography~\cite{benschop_extreme_2008,mizoguchi_first_2010,banine_physical_2011}. In these sources a liquid tin drop is impacted by a first laser pulse to shape the drop into a suitable target for plasma formation. A second pulse creates the plasma, where the line emission from excited tin ions provides the EUV light~\cite{banine_physical_2011,osullivan_spectroscopy_2015}. Our setup can be understood as large-scale model system to study the fluid dynamics of the target formation by the first laser impact. 
Figure~\ref{fig:ImpactRegimes} shows the response of a drop to a laser impact and how the outcome depends on the laser-pulse properties. The drop is accelerated as a whole by the localized phase change that consequently leads to a lateral expansion of the drop. For a tightly focused laser pulse (\Fig\ref{fig:ImpactRegimes}\,(a)) the drop breaks into tiny fragments during the expansion. By contrast, \Fig\ref{fig:ImpactRegimes}\,(b) shows the evolution of the drop into a curved liquid sheet that breaks at the edge of the sheet. The purpose of the apparatus is to quantify this response of the drop: starting with the laser impact, followed by the deformation of the drop and how it depends on the laser-pulse properties up to the final breakup of the liquid body into tiny fragments.
However, the features and capabilities of the setup described in this paper can be used to study laser-matter interactions and laser-induced liquid motion for a much broader range of experiments: the target, in our case a free-falling spherical drop, may be replaced by a planar geometry such as a liquid film to study the laser-induced forward motion, for example for non-Newtonian~\cite{wang_three-dimensional_2010} or high-viscosity~\cite{inui_laser-induced_2015} liquids, viscoelastic hydrogels as exploited for the printing of cells and biomaterials~\cite{guillemot_high-throughput_2010}, and even metals that are melted by the laser impact~\cite{visser_toward_2015}. In general, the laser impact upon a liquid target results in a forcing that is concentrated both in time and space, which can be visualized in our setup. The apparatus therefore allows for accurate and quantitative studies of laser-matter interactions that induce liquid motion.

We present in \Chap\ref{sec:Description} to \Chap\ref{sec:Control} an experimental setup that not only enables us to study but also to control the fluid-dynamic response of a target driven by a laser-induced vaporization. The scale of the experiment is set by the initial radius~$\Ro \approx 1\,\mathrm{mm}$ of a free-falling drop that is hit by a laser pulse emitted with a duration of $\taup=5\,\mathrm{ns}$ at a wavelength of $\lLaser=532\,\mathrm{nm}$.\@ We control the deposition of laser energy in the liquid by adding a dye such that the linear absorption coefficient~$\alpha$ at the wavelength~$\lLaser$ can be varied in comparison to \Ro\ over several orders of magnitude, i.e.~$5\times 10^{-5} \le \Ro\,\alpha \le 400$.
The spatial distribution of laser energy can further be tuned by the focusing condition of the laser in relation to the drop. We explain in detail how the energy distribution can be visualized and changed, either by a beam-shaping technique or varying $\alpha$. The purpose of our setup is to study the fluid-dynamic response of the liquid target to a laser impact. To this end, the experimental apparatus allows for extensive visualization of the complete process by high-speed and stroboscopic imaging. Having introduced the experimental apparatus, we present a few results of the system in \Chap\ref{sec:Results}: the stability of the control parameters of our experiment is discussed, followed by a brief introduction to the laser impact on a drop based on the stroboscopic imaging. Finally, we use the high-speed imaging for experiments showing how the fluid dynamics can be controlled by the laser-beam profile. We then give a summary of our work in \Chap\ref{sec:Summary}.

\section{System overview}\label{sec:Description}
A key feature of the experimental apparatus is the ability to control the laser impact in terms of the laser-pulse energy: both the absolute scale and the spatial distribution of energy can be controlled and visualized.
We present two optical configurations to impact the liquid target, which each have their own advantages.
First, a free beam propagation is introduced in \Chap\ref{sec:Laser} together with the overall layout of the setup (\Fig\ref{fig:Optics}, see Nomenclature for a list of abbreviations). This beam path leaves enough freedom to integrate other components such as additional optics to change the polarization state of the laser beam. In case more control over the spatial distribution of laser energy at the impact location is required we present a second configuration in \Chap\ref{sec:Near-fieldImaging}. This optical path incorporates a beam-shaping technique to first modify the beam profile in the near-field of the laser. Then, an imaging technique is applied to propagate the near-field image to the impact location where the drop is placed.

Equally important is the next aspect of our experiment: the visualization of the fluid-dynamic response of the drop to the laser impact. We describe in \Chap\ref{sec:Visualization} a combination of high-speed and stroboscopic imaging techniques that are incorporated in the laser-beam path. The response of the drop to the laser impact is then visualized in two orthogonal views: a side-view perpendicular to the laser beam and a back-view that is along the laser-beam propagation.

The spatial scale of the experiment is set by the initial radius \Ro\ of the drop, where we chose for a millimeter-sized drop based on three considerations. 
First, the ease to create drops of that size for common liquids as described in \Chap\ref{sec:Chamber}. Second, for this drop size the timescale of the fluid dynamics, or at least the late-time dynamics, is accessible by high-speed imaging, see \Chap\ref{sec:Visualization}. Third, stroboscopic imaging at this scale allows for recordings with high spatial and temporal resolution. In this case, the repetition rate of the experiment is set to~$\fexp=1\,\mathrm{Hz}$, limited by the frame rate of the high-resolution cameras used during stroboscopic recordings.

Preliminary experiments showed that to propel a millimeter-sized drop a laser pulse with an energy of a few $100\,\mathrm{mJ}$ is required. Commonly available at such pulse energies are Nd:YAG lasers with wavelength of $\lLaser = 532\,\mathrm{nm}$. However, to tune the linear absorption of light at that wavelength, i.e.\ to control the length scale on which the laser energy is absorbed in common liquids such as water, the addition of a dye is necessary as explained in \Chap\ref{sec:Liquids}. A high degree of control over the fluid dynamics and laser-impact conditions, required especially for stroboscopic imaging techniques, goes along with a precise control system, which is described in~\Chap\ref{sec:Control}.

\section{Laser system and free beam propagation}\label{sec:Laser}
\begin{figure}
	\centering 
	\includegraphics[width=\columnwidth]{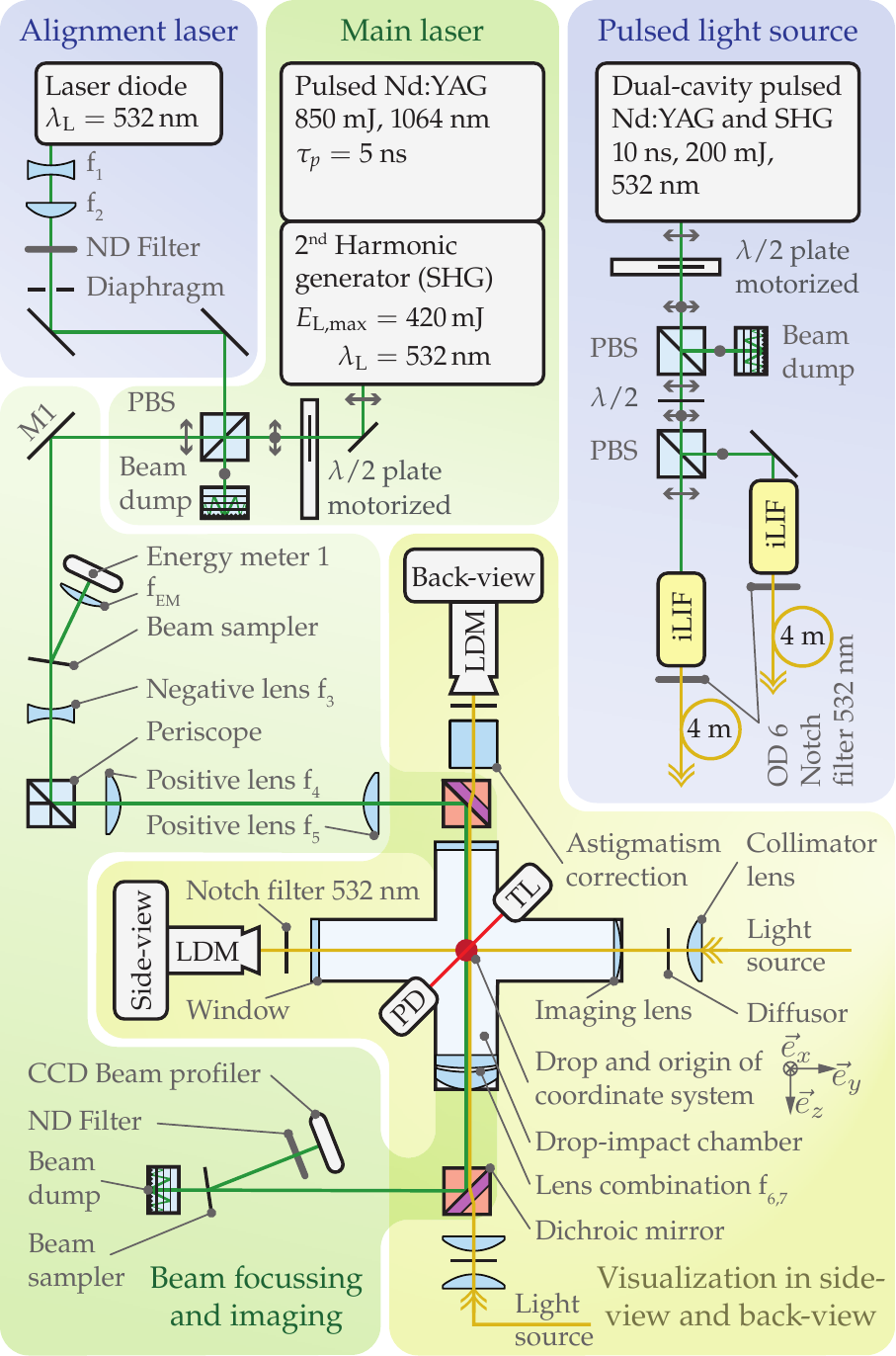}
	\caption{Sketch of the optical path of the laser system and visualization as arranged on an optical table with a size of $2.4\,\mathrm{m}$ x $1.5\,\mathrm{m}$ (shown not to scale). The symbols on top of the light path indicate out-of-plane (\protect\includegraphics{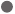}) and in-plane polarization (\protect\includegraphics{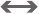}). Optical mirrors operated at an angle of incident (AOI) of $45^\circ$ are represented by black solid lines. The mirror labeled M1 serves as reference point in \Fig\ref{fig:BeamPath}, where an advanced laser-beam path is shown to scale. The drop defines the origin of our coordinate system, where $\vec{e}_z$ is aligned to the laser-beam propagation and $\vec{e}_x$ to gravity. Abbreviations that are used in the figure are explained in Nomenclature.\label{fig:Optics}} 
\end{figure}

The layout of the experimental apparatus with a free beam propagation is shown in \Fig\ref{fig:Optics}. The \emph{main laser} used to impact the liquid drop is a Q-switched Nd:YAG laser system (Q-smart 850 by Quantel) at its fundamental frequency. Its output is frequency-doubled in a temperature-stabilized crystal (second harmonic generator (SHG) by Quantel) to generate laser light at a wavelength of $\lLaser = 532\,\mathrm{nm}$ emitted within a pulse duration of $\taup=5\,\mathrm{ns}$ full width at half maximum (FWHM). The cavity optics and alignment are altered by the manufacturer to meet the design repetition rate of our experiment $\fexp = 1\,\mathrm{Hz}$ and delivers a slightly elliptical flat-top beam profile. The laser beam is linearly polarized, which allows for a convenient attenuation by splitting the beam in a polarizing beam splitter (PBS, PBS25-532-HP by Thorlabs or 2-HPCB-B-0254 by Altechna), see also \Fig\ref{fig:Optics}. The split-up ratio between the energy redirected into a beam dump and the remaining energy used in our impact experiment is set by a $\lambda/2$-plate in a motorized rotation mount (K10CR1/M by Thorlabs). That way we can set the energy \ELaser\ manually or by software control in the range $E_\mathrm{L,min} = 1\,\mathrm{mJ} \le \ELaser \le E_\mathrm{L,max} = 420\,\mathrm{mJ}$.

A continuous-wave laser diode (CPS532 by Thorlabs) serves as light source for the \emph{alignment laser}. It is coupled in the laser-beam path by the same PBS used for the attenuation of the main laser (\Fig\ref{fig:Optics}). The beam diameter of the diode laser is increased by a factor of three in a Galilean beam expander~\cite{weichel_laser_1990} ($\f{1} = -50\,\mathrm{mm}$, $\f{2} = 150\,\mathrm{mm}$) and can be matched precisely to the main laser by a diaphragm (see \Fig\ref{fig:Optics}). The alignment laser allows safe operation without laser-safety goggles when attenuated sufficiently by a neutral density (ND) filter.
The optical path of the main laser is not affected by the optical positioning of the alignment laser. Thereby, the total optical path length can be minimized, which is of particular importance when imaging the near-field as described in \Chap\ref{sec:Near-fieldImaging}.
In this section, we continue our description of a free beam propagation and general features such as the energy measurement, which are also used later in combination with the near-field imaging technique.

The laser energy \ELaser\ is measured by the energy meter~1 (energy sensor is a QE12LP-S-MB connected to an energy monitor S-LINK-2 by Gentec Electro-Optics). 
We use a wedged beam sampler (BSF10-A by Thorlabs) for in-plane polarized light to split-off 1\% by reflection at $45^\circ$ incidence. In case we change the polarization state, either to an out-of-plane or a mixed polarization state as will be done for the beam-shaping in \Chap\ref{sec:Near-fieldImaging}, we use a wedged window (WW11050-A by Thorlabs) with a dielectric coating as beam sampler at near-zero incidence as illustrated in \Fig\ref{fig:Optics} to split-off approximately 0.5\,\% of laser light, independent of the polarization state. A focusing lens $\f{\mathrm{EM}} = 150\,\mathrm{mm}$ matches the beam diameter to the sensor size of energy meter~1. For calibration purposes a second energy meter (same type as energy meter 1 but protected from the high-energy beam by an attenuator QED-12 by Gentec Electro-Optics) can be placed directly in the path of the main laser. This way the readout at the energy meter 1 can be calibrated against an arbitrary position along the laser-beam path. In particular, the dimensions of the sensor head allow for a direct placement at $z=0\,\mathrm{mm}$ within the drop-impact chamber (\Fig\ref{fig:Chamber}) to include any loss of light at optical elements in an appropriate calibration curve.

The laser beam is expanded in a Galilean beam expander ($\f{3} = -250\,\mathrm{mm}$, $\f{4} = 500\,\mathrm{mm}$), raised in height in a periscope assembly before it is focused by lens $\f{5} = 400\,\mathrm{mm}$ into the drop-impact chamber (see \Fig\ref{fig:Optics} and~\ref{fig:Chamber}). The beam expansion by a factor of $\f{4}/|\f{3}| = 2$ is required to increase the initial beam diameter $\Do=9\,\mathrm{mm}$ to a size that prevents the local fluence~$F$ to exceed the laser-induced damage threshold (LIDT) on any optical element. All optics have appropriate laser-line coatings for $\lLaser = 532\,\mathrm{nm}$ with a typical LIDT of $F_\mathrm{LIDT} = 5\,\mathrm{J}/\mathrm{cm}^2$ leading to a minimum spot size $d_\mathrm{min} \ge 2\, {(E_\mathrm{L,max}/(\pi\,F_\mathrm{LIDT}))}^{0.5} = 3.2\,\mathrm{mm}$ for each optical element at maximum laser energy. The periscope assembly allows us to adapt the height of the laser-beam path to about $250\,\mathrm{mm}$ above the optical table to accommodate the high-speed cameras that are placed on translational stages. Before the laser beam enters the chamber a dichroic mirror combines the laser-beam path and the optical path of the back-view visualization
(see~\Chap\ref{sec:Visualization}).

The diameter $d_0$ of the laser beam at the drop location can be adjusted by setting the position of lens \f{5} along the beam-propagation axis $\vec{e}_z$. The smallest beam diameter at the drop position is set by the beam waist $\omega_0$, which is of the order of a few tens of micrometer given our laser system and focusing conditions. This means that we can cover the range $ \omega_0/\Ro \approx 0 < d_0/(2\Ro) < 5$. In principle, even larger values are possible but out of scope from a fluid-dynamics point of view, since in such a case the laser fluence is too low to induce a considerable liquid motion.

Laser light that passes the drop is captured by a lens combination \f{6,7} to image the drop-impact location onto the charge-coupled device (CCD) of a beam profiler (BC106N-VIS/M by Thorlabs).\@ The combination of a plano-convex lens $\f{6} = 750\,\mathrm{mm}$ and a meniscus lens $\f{7} = 300\,\mathrm{mm}$ to a lens with an effective focal length~\cite{hecht_optics_2002} of $\f{6,7} = {(1/\f{6}+1/\f{7})}^{-1} = 214\,\mathrm{mm}$ is advantageous for two reasons. First, the flat surface of the plano-convex lens easily seals the drop-impact chamber against the ambient atmosphere (see \Fig\ref{fig:Chamber}).\@ Second, the combination reduces spherical aberrations and improves the image quality on the CCD.\@ To protect the delicate CCD from the high-energy beam, an attenuation by more than six orders of magnitude is required. The amount of attenuation must be insensitive to the polarization state of the light to capture all phase components at the same relative intensity. Otherwise, the sensor may only capture the beam profile for light with in-plane polarization, which may be different from the profile for the out-of-plane component (a corresponding polarization state occurs for the beam-shaping in \Chap\ref{sec:Near-fieldImaging}).\@ As first attenuation step, we choose a wedged window at near-zero incidence as beam sampler. The final step is a neutral density (ND) filter that can be exchanged easily to match the sensitivity of the CCD to a change of laser energy. The position of lens \f{6,7} on the optical path relative to the drop as well as the position of the CCD thereafter are fixed by the Gaussian lens formula~\cite{hecht_optics_2002}, which is explained in more detail in \Chap\ref{sec:Near-fieldImaging}.

For experiments that require laser energies much lower than $E_\mathrm{L,min}$ we introduce an additional step of attenuation not shown in \Fig\ref{fig:Optics}.\@ Two wedged windows can be placed between the beam sampler and lens \f{3} in a Z-con\-fi\-gu\-ra\-tion to use the reflection of the first face of each window. The coating on the windows can be chosen to reach an additional attenuation by a factor of $10^{4}$ for two coated windows or a factor of $10^{2}$ in case one window is replaced by a laser-line mirror. 

The free beam propagation described here is simple since it does not pose many restriction on the optical path. The distance between lens \f{3} and \f{4} is fixed to the sum of the focal length, $\f{3}+\f{4}$, to keep a collimated beam after the expansion. The position of lens \f{5} relative to the drop is crucial to set the focusing condition. This leaves enough freedom to integrate other components in the beam path such as a $\lambda/4$- or $\lambda/2$-plate to change the polarization state. In case more control over the beam shape is required we present an imaging and beam shaping technique in the following section.

\section{Beam shaping and near-field imaging}\label{sec:Near-fieldImaging}
A drawback of a free beam propagation as described in the previous section is the deterioration of the flat-top beam profile as it propagates from the near-field to the far-field at the drop-impact location ($z=0\,\mathrm{mm}$). This can be solved by imaging the flat-top profile from the near-field (at a distance of about $400\,\mathrm{mm}$ from the laser head) to the drop-impact location (at a distance of about $2500\,\mathrm{mm}$) as described in this section. In addition, the intrinsic properties of an imaging technique are beneficial when combined to a beam-shaping operation, which we explain in this section as well.

A near-field imaging is advantageous when a beam-shaping operation is performed at, or close to, the image plane that has its conjugated focal plane at the drop-impact location: the image is then propagated to its conjugated focal plane independent of the actual light path (of course, as long as clipping or spherical aberrations are minimized). As a consequence, a deviation in the propagation angle, inevitably introduced by the beam shaping, does not compromise the profile and, more importantly, the final position of the beam in the conjugated focal plane where the target of the laser impact is placed. As a result, once the imaging is implemented and properly aligned the beam shaping can be performed with minimum alignment effort. These considerations motivated the beam path presented here: a beam-shaping realized within the near-field distance such that any linear combination of the laser beam with a rotated version of itself can be imaged onto the drop. In addition, we place the CCD of the beam profiler in another conjugated focal plane behind the drop-impact location for the visualization of the beam shape as experienced by the drop (already explained briefly in \Chap\ref{sec:Laser}). Such a setup allows to study the response of the drop to different laser-beam profiles.

\begin{figure*}[t!]
	\begin{minipage}[t]{0.50\textwidth}
		\mbox{}\\
		\includegraphics[width=1.412\columnwidth]{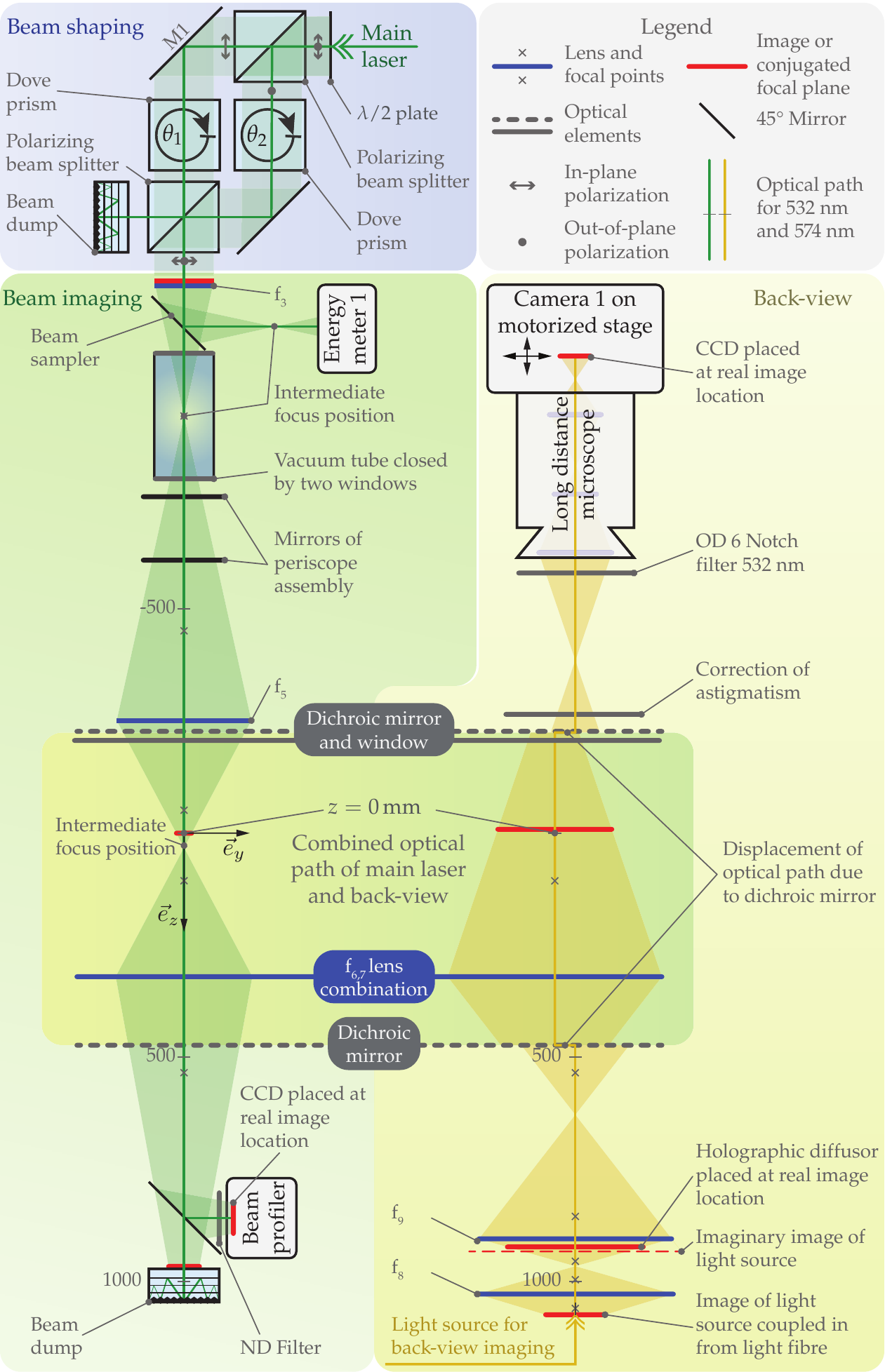}
		\end{minipage}\hfill
		\begin{minipage}[t]{0.26\textwidth}
			\mbox{}\\[-\baselineskip]
			\caption{Unfolded optical path for the main laser and back-view visualization. The drawings for the \emph{beam imaging} and \emph{back-view} are to scale and aligned, but make use of two different scales in $\vec{e}_z$- and $\vec{e}_y$-direction to accommodate the complete beam path of about $2500\,\mathrm{mm}$ in length, while the maximum lateral extension in $\vec{e}_y$ is only $30\,\mathrm{mm}$.\@ As a consequence, the split-off at energy meter 1 and the beam profiler are shown at $45^\circ$ AOI for illustration, whereas they are actually performed at near-zero AOI.\@ Similarly, the reflections at $45^\circ$ AOI at any mirror are omitted in this unfolded representation of the optical path (mirror M1 is marked as reference point to compare with \Fig\ref{fig:Optics}).
			The direction of light propagation of the back-view is reversed in comparison to the main laser to avoid forward scattering or even direct laser radiation into the high-speed camera in case any protective measure such as a notch filter fails. The imaging of the near-field beam profile of the laser onto the drop is achieved by lens \f{3} and \f{5} as explained in the text.\label{fig:BeamPath}}
		\end{minipage}
	\end{figure*}

The complete laser-beam path including the \emph{beam shaping} is shown in \Fig\ref{fig:BeamPath}.\@ As before, we make use of the polarization state of the main laser and create two orthogonal fields by a $\lambda/2$-plate and a PBS.\@ Two dove prisms allow for an arbitrary image rotation $\theta$ of the two fields. The multiple reflections inside the dove prism cause a slight change of the polarization state of each beam, which leads to a loss of laser energy of about $10\%$ when the two fields are recombined in another PBS.\@ Such an arrangement allows to create any linear combination of the input beam profile with a rotated version of itself. We use this feature to transform the elliptical beam profile to a near-circular profile to achieve axisymmetric impact conditions on the drop. The complete arrangement including attenuation as described in \Chap\ref{sec:Laser} is achieved within the near-field of the laser system (about $400\,\mathrm{mm}$).\@

\begin{figure*}[t!]
	\begin{minipage}[t]{0.50\textwidth}
		\mbox{}\\
		\includegraphics{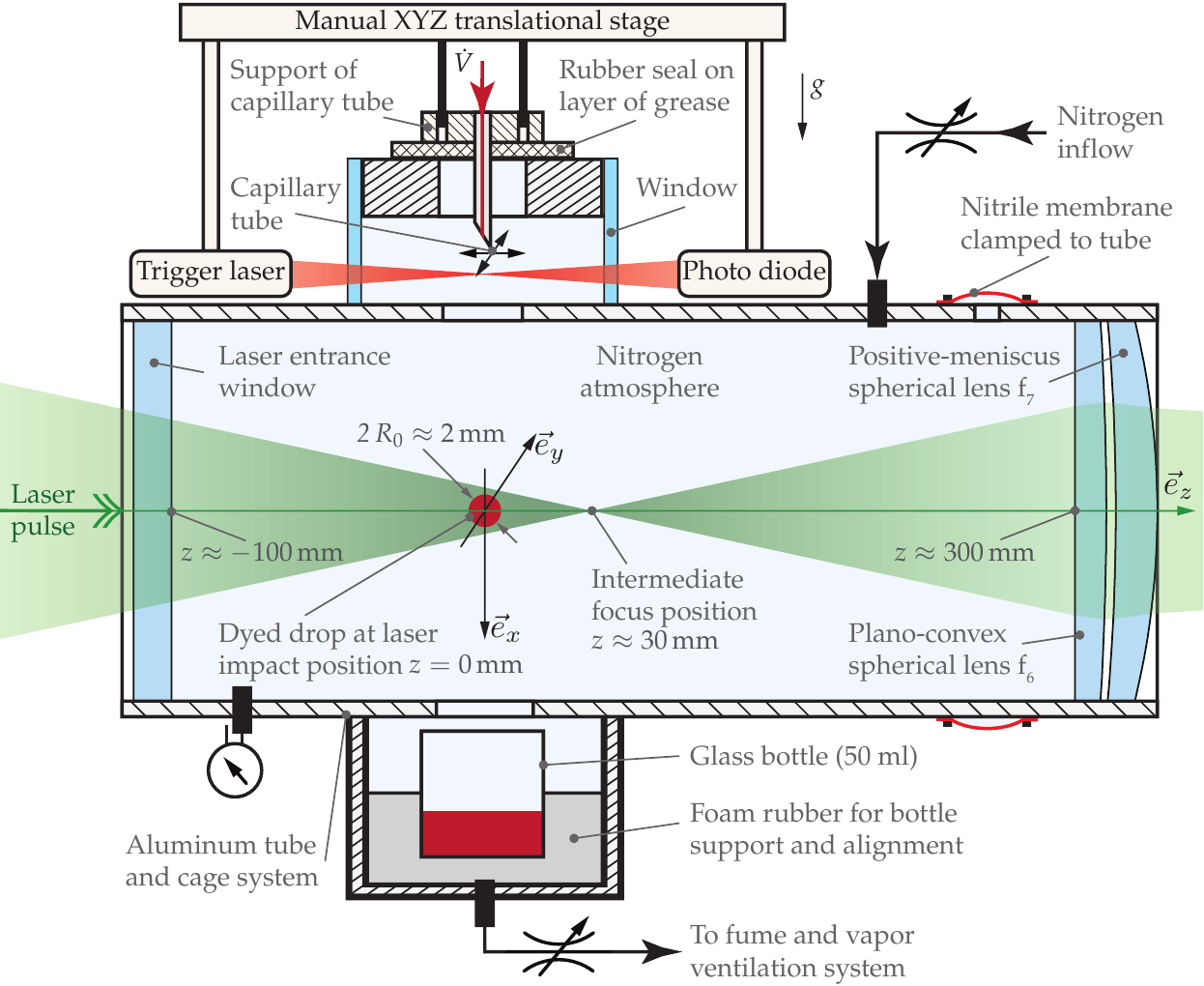}
		\end{minipage}\hfill
		\begin{minipage}[t]{0.26\textwidth}
			\mbox{}\\[-\baselineskip]
			\caption{A sketch of the drop-impact chamber in a cross-section illustrating the moment of laser impact (dimensions not to scale, intermediate focus position as calculated in \Chap\ref{sec:Near-fieldImaging}). The inner dimensions of the aluminum chamber are such that common 50-mm and 2-in.\ optics can be mounted directly. The chamber is sealed against ambient air and multiple nitrogen inlet ports are placed close to each optical element (illustrated for the lens combination at the right). A single outlet at the bottom of the chamber is connected to a low-pressure ventilation system leading to gentle flow away from the optical elements to the bottom of the chamber. Any excessive liquid that does not evaporate over time is collected at the bottom of the chamber in a glass bottle that allows for a quick exchange.\label{fig:Chamber}}
		\end{minipage}
	\end{figure*}

In principle, the beam can be left freely propagating from this position (marked by the position of lens \f{3} in \Fig\ref{fig:BeamPath}) to the drop-impact location as described in \Chap\ref{sec:Laser}. But as discussed before, any deviation in angle inevitably introduced by the dove prisms would lead to a large offset at the drop-impact location due to the long optical path of about $2500\,\mathrm{mm}$. This would need a careful re-alignment for each rotation of the beam, which we avoid by the near-field imaging. However, a beam-imaging technique imposes several restrictions upon our setup:
\begin{enumerate}
	\item Any intermediate focus point of the high-energy laser beam needs to be covered with a vacuum tube to avoid optical breakdown in ambient air before the laser pulse hits the drop, i.e.\ for $z<0\,\mathrm{mm}$. Otherwise, the plasma at the focus spot absorbs and scatters nearly 100\,\% of the incident laser energy. 
	\item No optical element can be placed in the vicinity of the drop to avoid the impact of fragments onto optical elements causing a damage or alteration to the laser-beam path in subsequent experiments.
	\item The damage threshold $F_\mathrm{LIDT}$ for each optical element must be respected as explained before in~\Chap\ref{sec:Laser}.
	\item Two dichroic mirrors with a diameter of $50\,\mathrm{mm}$ at $45^\circ$ incidence need to be incorporated to allow sufficient access for the back-view visualization.
	\item The optical elements placed at the laser entrance and exit of the drop-impact chamber need to allow for a proper seal of the inert atmosphere (compare \Fig\ref{fig:Chamber}).
	\item In order to visualize the high-energy beam on the beam profiler we must attenuate the light once transmitted through the chamber. When the beam shaping is to be used, the laser beam contains an in-plane and out-of-plane polarization state and the attenuation must be performed equally for all phase components. We described a possible technique, a reflection at $0^\circ$ incidence, already in \Chap\ref{sec:Laser}, but its implications on the path length must be considered here as well.
\end{enumerate}

We design the laser path by an iterative procedure using geometrical optics, which is justified as the length scales to be considered here are large compared to the wavelength of the laser, that is $\Ro/\lLaser \gg 1$. Each imaging step is then described by the Gaussian lens formula~\cite{hecht_optics_2002},
\begin{eqnarray}
	\frac{1}{s_i} + \frac{1}{s_i'} &=& \frac{1}{\f{i}}\,,\label{eq:Imaging}\\
	M_i &=& \frac{s_i'}{s_i} = \frac{d_i'}{d_i}, \label{eq:Magnification}
\end{eqnarray}
that relates the distances from the object to the lens $s_i$ and the distance from the lens to the image $s_i'$ to the focal length \f{i}\ and magnification $M_i$ of the imaging step $i$.\@ The above equations need to satisfy the following restrictions: the magnification of the near-field from the first imaging step to the drop $M_\mathrm{drop} = \prod^\mathrm{drop}_1 M_i \ge 2\,\Ro/\Do = 0.25$, in order to illuminate the drop completely. Likewise, the magnification from the drop to the CCD of the beam profiler $M_\mathrm{BP} = \prod_\mathrm{drop}^\mathrm{BP} M_i \le d_\mathrm{BP}/{(\Do\, M_\mathrm{drop} )} \approx 2.9$ to visualize the complete beam on the CCD.\@

As the required magnification to match the beam size in the near-field to the drop size satisfies $M_\mathrm{drop} < 1$, the occurrence of an intermediate focus point would inevitably be close to the conjugated focal plane if a collimated laser beam was imaged (the position of the intermediate focus point would be at the focal point of the imaging lens at $z<0\,\mathrm{mm}$). We then recognize that at least one additional imaging step is required to create a diverging beam such that the intermediate focus point is shifted to a position $z>0\,\mathrm{mm}$ while the conjugated focal plane is kept at at the drop position, $z=0\,\mathrm{mm}$. Despite those considerations and restrictions, the system is still underdetermined given the choices of optical elements available. Therefore, we first pick a set of optical elements, choose $s_1$ and calculate any remaining quantities according to Eqs.~(\ref{eq:Imaging}) and (\ref{eq:Magnification}). We calculate the local fluence at all optical elements to evaluate our choice based on restrictions 1 to 6. Next, we determine and evaluate the consequences for the back-view visualization. 

The illumination for the back-view visualization needs to be partially incorporated in the optical path of the main laser given the position of the dichroic mirrors. We calculate the image of the light source along its optical path based on Eqs.~(\ref{eq:Imaging}) and (\ref{eq:Magnification}). However, the optical elements that are part of the combined optical path of the main laser and back-view visualization are already fixed in terms of their position and focal length. In addition, the divergent nature of the light source requires the first lens in the back-view path to be a lens (or focusing mirror) with a large numerical aperture ($\mathit{NA}$) to capture most light of the source (we select an aspheric collimator lens $\f{8}=32\,\mathrm{mm}$). This narrows down the choice for the set of lenses already considerably. The remaining criterion to check for the back-view illumination is then given by its sole purpose: the required magnification to illuminate the drop-impact position in a uniform way. As will be discussed in \Chap\ref{sec:Visualization} the size of the field of view (FOV) $d_\mathrm{FOV} = 20\,\Ro =20\,\mathrm{mm}$ and given the lateral extension of the light source, typically about $d_\mathrm{Light} = 8\,\mathrm{mm}$, we find for the required magnification $M_\mathrm{BV} = d_\mathrm{FOV} / d_\mathrm{Light} \ge 2.5$.

Our solution to above problem is illustrated in \Fig\ref{fig:BeamPath} with one imaging step for the laser-beam path before (lens \f{3} and \f{5}) and one step after the drop (lens \f{6} and \f{7}). The back-view illumination includes two imaging steps before the drop, which is necessary to compensate for the effect of the shared lens $\f{6,7}=214\,\mathrm{mm}$ in the back-view by lens $\f{9}=50\,\mathrm{mm}$. The image of the laser beam at the position of lens $\f{3}=300\,\mathrm{mm}$ is propagated to the drop position (image and conjugated focal planes are highlighted in red in \Fig\ref{fig:BeamPath}). These arrangements leads to the first intermediate focus after lens \f{3}, which is covered by a vacuum tube. As required, the next image of the laser-beam profile created by lens $\f{5} = 200\,\mathrm{mm}$ at the drop-impact position is located before the second intermediate focus at about $z=30\,\mathrm{mm}$ (see also \Fig\ref{fig:Chamber}). Thus, no laser energy is lost in an intermediate focus by plasma formation before the impact onto the drop.

The focus point behind the drop, i.e.\ the point of optical breakdown and plasma formation, cannot be covered by a vacuum tube. However, the distance of about $30\,\mathrm{mm}$ between the drop and the focus point is sufficiently large in order not to interfere with our side- and back-view visualization. Care must be taken to measure the beam profile, which is obtained in the absence of the drop by the beam profiler (\Fig\ref{fig:BeamPath}), at energies low enough to avoid optical breakdown.
Please note, that the intermediate focus just before energy meter 1 does not require a vacuum enclosure as the energy in that part of the laser-beam path is far below the threshold for optical breakdown. Therefore, lens \f{3} is not only used to image the near-field but also used to match the laser beam to the size of the sensor for energy meter 1.

Figure~\ref{fig:BeamPath} also illustrates the point that limits the maximum size of the FOV in the back-view: at the position of lens \f{6,7} the path of the back-view exhibits its maximum lateral extension. Reversing the order of lens \f{6,7} and the dichroic mirror could solve this issue but requires an additional window to seal the chamber before the dichroic mirror can be placed. This procedure would make the dichroic mirror the element that clips the FOV leading to no further improvement. Using lens \f{6,7} to seal the chamber as sketched in \Fig\ref{fig:Chamber} minimizes the required path length and is thus our preferred choice. In this configuration, the light enters the \emph{back-view} at the bottom of \Fig\ref{fig:BeamPath} and forms a real image after lens \f{8}. This image is propagated by lens \f{9} and \f{6,7} to the drop impact location with the appropriate magnification. The light source is divergent as illustrated in \Fig\ref{fig:BeamPath} and light is lost in the additional imaging step. Therefore, a diffusor with fine control over diffusing angles is required (holographic diffusers by Edmund Optics, version with $5^\circ$ to $20^\circ$ diffusing angle). This selection allows to find a compromise between light intensity and uniformity. Sufficient light must be captured by each camera as described in \Chap\ref{sec:Visualization} while the FOV needs to be illuminated uniformly to ease subsequent image analysis and interpretation.

\section{Visualization}\label{sec:Visualization}
The purpose of our setup is to study the fluid-dynamic response of the liquid target to a laser impact. To this end, the experimental apparatus allows for extensive visualization of the complete process by high-speed and stroboscopic imaging as explained in this section. The response of the liquid drop to the laser impact involves several time\-scales. The laser impact itself takes place on a nanosecond timescale set by $\taup = 5\,\mathrm{ns}$. The timescale of the drop dynamics ranges from a few microseconds (compare \Fig\ref{fig:ImpactRegimes}) up to the capillary timescale $\tauc = {(\rho \Ro^3/\gamma)}^{1/2} \approx 1\,\mathrm{ms}$,
where $\rho$ and $\gamma$ are the liquid density and surface tension, respectively. The frame rate $f_\mathrm{cam}$ to resolve the largest timescale $\tauc$ properly is at least $5000$ frames per second (fps) following typical criteria for high-speed imaging~\cite{versluis_high-speed_2013}. Any other process in our experiment requires a frame rate of orders of magnitude larger.

The spatial scale of interest is set by $\Ro$ but needs to allow for an expansion of the initially spherical drop into a larger geometry, where preliminary experiments suggested a field of view (FOV) of about $d_\mathrm{FOV} = 20\,\Ro=20\,\mathrm{mm}$ in diameter. However, the smallest features to be resolved are those of drop fragments $\ll \Ro$, see \Fig\ref{fig:ImpactRegimes}. This requires an imaging technique with a high spatial resolution to capture both the large FOV and the details of drop fragmentation. Furthermore, the liquid target is allowed to move freely in space after the laser impact, which requires that no optical element is present in the vicinity of the drop to avoid liquid fragments reaching the optics and causing a damage or alteration to the laser-beam path in subsequent experiments. Since the laser impact can be violent as shown in \Fig\ref{fig:ImpactRegimes}, a distance from the drop of about $100\,\Ro$ to $300\,\Ro$ for each optical element is preferred. Thus, imaging at a large working distance in comparison to $d_\mathrm{FOV}$ is required.

Our visualization techniques must allow for an optimum temporal and spatial resolution given the aforementioned considerations, which is difficult to meet with a single system. Therefore, we choose for high-speed and stroboscopic imaging~\cite{versluis_high-speed_2013} techniques to be used interchangeable in our setup. The latter requires an experiment with a high degree of reproducibility to allow for multiple experiments being recorded under identical conditions but imaged at different times to give the perception of a continuous movie.

To image the laser impact and drop onto the different cameras we use a long-distance microscope (LDM, K2 DistaMax by Infinity Photo-Optical Company) for both high-speed and stroboscopic imaging. The LDM supports full-frame camera chip sizes ($36\,\mathrm{mm}$ x $24\,\mathrm{mm}$) as found in the high-re\-so\-lu\-tion cameras ($4008 \times 2672 \sim 10^7$ pixels) that we use for stroboscopic imaging. The high-speed cameras feature smaller chip sizes ($20\,\mathrm{mm}$ x $20\,\mathrm{mm}$, $1024 \times 1024 \sim 10^6$ pixels) that require the magnification to be adapted by exchangeable objectives. Fortunately, the magnification required in our case, given the FOV and camera chip sizes, is of the order of one, which does not pose great demands on the imaging optics except the large working distance.

The high-speed system consists of a continuous light source (LS-M352A metal halide light source by SUMITA Optical Glass for the back-view and a MAX-303 xenon light source by Asahi Spectra for the side-view) that is prepared by critical illumination (see \Fig\ref{fig:Optics}). An aspheric collimator lens (ACL5040U-A by Thorlabs) creates an image of the light source on a diffusor with a high transmission efficiency (ED1-C50 by Thorlabs and holographic diffusers by Edmund Optics). A second aspheric lens (AL50100-A) projects a magnified version of that image to fill the FOV of the side-view in a uniform way (the optical path for the back-view is described in \Chap\ref{sec:Near-fieldImaging}). To allow for a precise alignment the optics for each light source are mounted on a yaw-and-pitch platform on top of a translational stage. The high-speed cameras (FASTCAM SA-X2 and FASTCAM SA1.1 by Photron for the back- and side-view, respectively) are mounted on translational stages. The one for the back-view is motorized to move the image plane along the optical axis $\vec{e}_z$ by software control.
This feature allows to keep the drop dynamics in the focus of the visualization:
 the drop is propelled along $\vec{e}_z$ and, therefore, leaves the depth of field at some point in case the camera position is fixed, even at nearly-closed aperture of the LDM.\@ 

The back-view poses another challenge: it needs to be combined with the laser-beam path as already discussed in \Chap\ref{sec:Near-fieldImaging}. This is accomplished by two custom-made dichroic mirrors (Laseroptik, Garbsen). These long-pass filter are phase preserving to keep the polarization state, have a high $F_\mathrm{LIDT}$ and a near-100\% transmission for $\lambda \ge 574\,\mathrm{nm}$. The latter is a requirement for the monochromatic light source that we use for stroboscopic imaging.

The light path of the back-view and its alignment in reverse direction with the laser-beam path is shown to scale in \Fig\ref{fig:BeamPath}. The direction of propagation is reversed to avoid forward scattering or even direct laser radiation into the high-speed camera in case any protective measure such as a notch filter fails. The thickness of the glass substrate of the dichroic mirrors introduces a considerable displacement of the optical path (\Fig\ref{fig:BeamPath}), which leads to an astigmatism in the back-view that has to be corrected for proper imaging. A substrate of the same material and dimensions but with an anti-reflective coating is introduced just behind the dichroic mirror rotated by $90^\circ$ about $\vec{e}_z$. This element introduces a displacement in $\vec{e}_x$-direction (not visible in \Fig\ref{fig:BeamPath}) to compensate the astigmatism by lengthening the optical path in the direction orthogonal to the displacement introduced by the dichroic mirror. Despite the additional optics and with careful alignment the amount of light captured by the high-speed cameras is enough to use even the shortest possible exposure time of $293\,\mathrm{ns}$ for the FASTCAM SA-X2. This option is very advantageous for our problem that involves multiple timescales. The short exposure time allows us to image drop fragments resulting from the breakup of the liquid drop without motion blur despite their small size and high speed, which clearly is beyond a few tens of meters per second.

The purpose of the stroboscopic visualization is to achieve far higher temporal and spatial resolution compared to high-speed imaging. We use pulsed light sources with exposure times of a few nanoseconds to illuminate the scene of interest and capture the image on high-resolution cameras. Both cameras (pco.4000 by PCO AG and Bigeye G-1100B Cool by Allied Vision Technologies) feature the same chip (Truesense KAI-11002 with $4008 \times 2672 \sim 10^7$ pixels) and mounting options (F-mount by Nikon), which allows identical LDM configurations to be used in side-view and in back-view. 
The resolution of the imaging system is estimated under actual experimental conditions (working distance of 320 to $400\,\mathrm{mm}$) based on images of a variable line grating test target (R1L3S6P by Thorlabs, placed at the drop impact position) to be within 26 and 50 line pairs per millimeter. The pco.4000 offers a double-image mode that allows to capture two successive frames of the same experiment with an inter-frame time of less than $1\,\upmu\mathrm{s}$. This feature can be used to quantify velocity maps, e.g.\ particle tracking and particle imaging velocimetry~\cite{versluis_high-speed_2013}. Of course, it also requires a pulsed light source capable of delivering two pulses of light in rapid succession. 

We use a dual-cavity, Q-switched, and frequency-doubled Nd:YAG laser system (EverGreen 200 by Quantel) as light source for stroboscopic imaging that is capable of delivering two pulses of laser light with a duration of $10\,\mathrm{ns}$ and an arbitrary delay between the two pulses (see \Fig\ref{fig:Optics}). The green laser is first attenuated based on its polarization in the same way as the main laser described in \Chap\ref{sec:Laser}.\@ Next, the beam is split in two parts by another $\lambda/2$-plate and a PBS.\@ The coherence of the laser light, which would lead to interference and speckles when used directly for visualization, is removed by incoherent laser-induced fluorescence light illumination (iLIF)~\cite{bos_ilif:_2011}.\@ In this technique, the laser beam is expanded into a dye solution or solid plate where the incoming light at a wavelength of $532\,\mathrm{nm}$ is absorbed to a resonantly excited state. The subsequent fluorescence emission at a wavelength of $574$ to $580\,\mathrm{nm}$ preserves the 10-ns pulse duration of the incoming laser pulse (the iLIF system coupled to the laser system is commercially available as High Efficiency Diffuser by LaVision).\@ The incoherent, monochromatic light is coupled into two optical fibers to allow an arbitrary placement of the light sources in the setup for imaging. Since the two pulses of light originate from a single laser pulse and differences in optical path length are minimized, the images taken with this light source are intrinsically synchronized in time.

All visualization systems described so far record grayscale images chosen for its superior light sensitivity~\cite{versluis_high-speed_2013}. A system to record color information requires light of a broad spectrum for illumination, which is challenging for the back-view as the dichroic mirrors act as a long-pass filter. However, the side-view only requires a notch filter, which minimizes the effect on the spectrum and offers a possibility to record color images. We use a consumer digital single-lens reflex (SLR) camera (for practical reasons, a Nikon D5100 is chosen for the images in \Fig\ref{fig:ImpactRegimes}, but for higher image quality a full-frame and low-noise version such as a Nikon D810 could be considered) to record color images stroboscopically and a pulsed light source (NANOLITE KL-K by HSPS) for flash illumination. The light source delivers a high-intensity light pulse of $8\,\mathrm{ns}$ in duration. It makes use of an electric discharge to create a plasma spark, which we image in a similar way as we described for the monochromatic light sources. A valuable extension for the sideview visualization is a Schlieren technique~\cite{vogel_sensitive_2006}, which would enable to visualize density gradients in the gaseous phase due to local vaporization or shock waves induced by the laser impact.

The images taken by the camera systems and beam profiler are calibrated by taking images of semi-transparent calibration targets placed at the drop-impact location. We use high-precision Ronchi rulings by Edmund Optics that cover the complete FOV or fixed-frequency grid distortion targets by Thorlabs. These targets allow to translate image dimensions in each view to world units in the lab reference frame. The spatial relation among the different views can be inferred either from a unique point on the calibration targets captured in all views at the same time or from the spherical drop itself. The initial drop before laser impact is recorded in all views, sets the origin for our coordinate system, and is therefore a natural reference point for calibration.

\section{Drop impact chamber}\label{sec:Chamber} 
The chamber encloses the laser impact in an inert atmosphere and limits the area where any fragments, vapor, or aerosol of the liquid may go after the laser impact. The chamber is built from a standard 60-mm cage system by Thorlabs with custom alterations to seal the aluminum housing and connect flow in- and outlet ports, see \Fig\ref{fig:Chamber}. The dimensions of the chamber are such that common $50$-mm or 2-in.\ optics can be mounted directly. The distances between the drop and the first optical element along the $\vec{e}_y$- and $\vec{e}_z$-axis are maximized under the restriction imposed by the near-field imaging and visualization. This requirement leads to the cross-shaped chamber design as seen in the top view in \Fig\ref{fig:Optics}.

The drop generator is found on top of the chamber in \Fig\ref{fig:Chamber} and fed by a constant flow rate $\dot{V}$ driven by a syringe pump (PHD Ultra by Harvard Apparatus, not shown). The design repetition rate $\fexp$ requires $\dot{V}=4/3\,\pi\, \Ro^3 \, \fexp = 0.25\, \mathrm{ml/min}$ for a drop of initial radius $\Ro = 1\,\mathrm{mm}$. A suitable principle for drop generation in this case is the quasi-static pinch-off at the tip of a capillary tube. The tube's outer radius $R_{c,o}$ needs to be smaller than the drop for a reproducible position of the drop detachment under the influence of gravity $g$. The Weber number $\We_c$\ of the capillary tube relates the kinetic energy of the liquid to its surface energy and can be expressed as $\We_c = \rho \Ro \dot{V}^2/(\pi^2 \gamma R_{c,i}^4)$, where the kinetic energy is determined by the flow speed inside the capillary with inner radius $R_{c,i}$. A regular 30-gauge needle with $R_{c,o}/\Ro = 1/3$ gives $\We_c = 0.6 < 1$, which means that control over the pinch-off position can be ensured for such a small needle while the drop generation is purely controlled by surface-tension forces and gravity. Any dripping, or even jetting~\cite{clanet_transition_1999, eggers_physics_2008} at $\We_c \gg 1$, is only observed in the case that $R_c$ is further decreased, for example in the event of clogging inside the tube.

A drawback of the proposed method is that it is impossible to vary the drop radius \Ro\ over a large range. \Ro\ is set by the balance of surface tension and gravity and scales as $\Ro \sim \kappa^{-2/3}\, R_{c,o}^{1/3}$, where $\kappa^{-1} = {(\gamma/{(\rho g)})}^{1/2}$ is the capillary length~\cite{de_gennes_capillarity_2002}. The weak dependence of $\Ro$ on $R_{c,o}$ does not allow for a large variation, but the drop generation mechanism is very robust and leads to a stable $\Ro \approx 1\,\mathrm{mm}$ for the liquids described in \Chap\ref{sec:Liquids}. 

Once the drop detaches at the tip of the capillary tube, it falls down towards the laser-impact position under the influence of gravity while it relaxes to a spherical shape. The drop masks a photodiode (PDA36A by Thorlabs) that is illuminated by continuous-wave laser diode (CPS635R by Thorlabs) emitting light at $635\,\mathrm{nm}$. The low-power laser diode is focused at the position where the drop passes as illustrated in \Fig\ref{fig:Chamber} to create a precise trigger on the passage of the drop.
The trigger finally leads to the main laser emitting a pulse of light that enters from the left through a window, hits the drop at $z=0\,\mathrm{m}$, and exits to the right through the lens combination \f{6,7}. The complete arrangement of the trigger laser, photodiode, and capillary can be moved relative to the drop chamber allowing for a free positioning of the drop along $\vec{e}_y$ and $\vec{e}_z$ without the need to change the trigger alignment. The details on the timing and control, used to align the drop relative to the laser beam along $\vec{e}_x$, are left for \Chap\ref{sec:Control}.

An elastic membrane visible from the outside of the chamber as shown in \Fig\ref{fig:Chamber} serves two purposes. First, it is a visual indication of a slight over-pressure inside the chamber relative to ambient conditions. The over-pressure is controlled by two flow-vales and a static pressure regulator (not shown) and ensures that no oxygen enters the chamber to prevent an explosive mixture with the vapor of a flammable liquid. Second, the membrane serves as safety release valve when an unexpected pressure build-up occurs during an explosion of flammable liquids that could not be prevented by other means. As discussed, the intermediate focus position at $z >0\,\mathrm{mm}$ cannot be covered by a vacuum tube and is a likely spot for the ignition of an explosion as previous experiments have confirmed. The over-pressure can also be checked with a digital differential pressure gauge, that offers an audible feedback in case the over-pressure is lost.

\section{Liquids}\label{sec:Liquids}
We aim to control the deposition of laser energy in the liquid by adding a dye such that the linear absorption coefficient~$\alpha$ at the wavelength~$\lLaser$ can be varied in comparison to \Ro. Water is a convenient model system in terms of its well-known physical properties, safe operation, availability, and previous research on laser-matter interaction~\cite{vogel_mechanisms_2003}. However, the linear absorption coefficient~\cite{pope_absorption_1997} $\alpha = 0.0447\,\mathrm{m}^{-1}$ of pure water at $\lLaser=532\,\mathrm{nm}$ is negligible small in comparison to the scale of our system, i.e.\ $\Ro\,\alpha \ll 1$. The opposite regime, $\Ro\,\alpha \gg 1$, is of great interest, since the laser energy is then limited to a superficial layer of the drop, the thickness of which scales as~$\delta \sim 1/\alpha$, which leads to a high energy density upon laser impact.
As a result, the kinetic energy imparted to the drop is increased~\cite{klein_drop_2015} and the fluid-dynamic response is more violent.
To achieve local absorption and actually control the penetration depth of laser light~$\delta$ at $\lLaser=532\,\mathrm{nm}$ into the liquid phase, we solve a dye at variable concentration in otherwise pure water.

To explore the range $\Ro\,\alpha \gg 1$ a dye with a high solubility is required. We use the dye Acid-Red-1 (chemical abstracts service registry number (CASRN): 3734-67-6, product number 210633 by Sigma-Aldrich), which is also used in coatings or inks and as such is available at large quantities and at reasonable cost. We filter each dye-water solution during preparation and place a syringe filter with a pore size of $200\,\mathrm{nm}$ in front of the capillary tube as final stage of filtration. This is of importance when performing experiments with a solution close to the solubility limit to ensure that no particles or dye agglomerates change the dynamic response of the system. This potential effect of particles in the liquid on the fluid dynamics is also why we avoid to use pigment dyes. Instead, we may use black (IJC-5900) and magenta (IJC-5920) pigment-free inks by Sensient Imaging Technologies.

\begin{figure}
	\centering
	\includegraphics{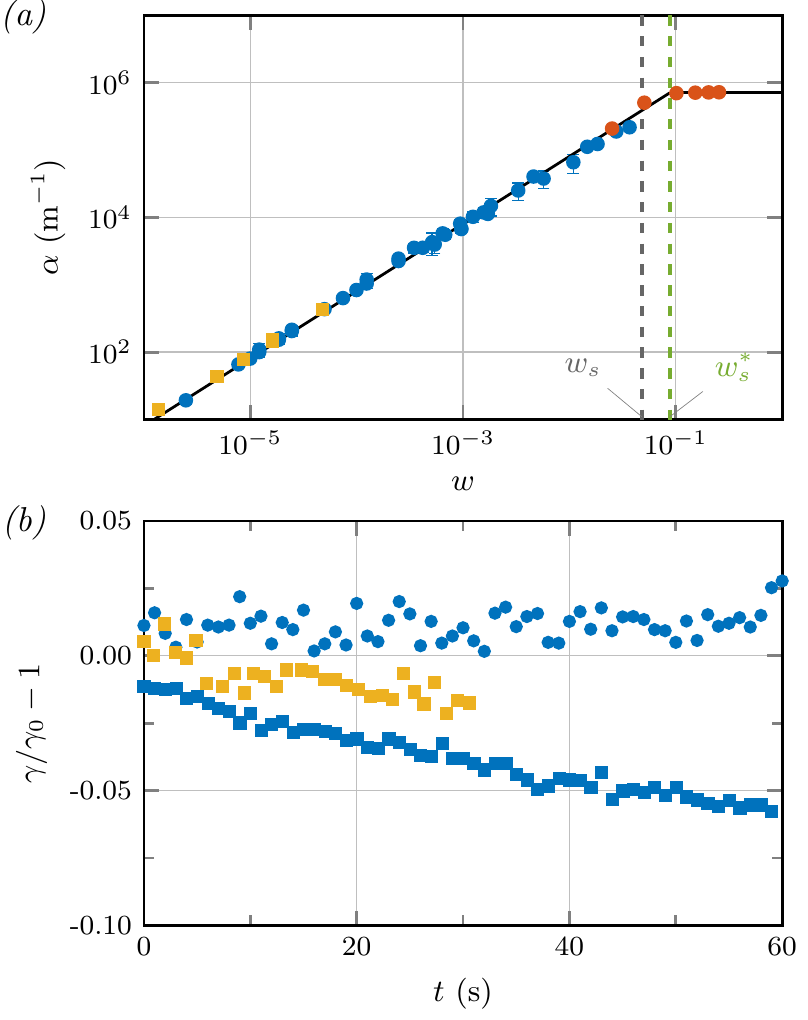}
	\caption{Liquid properties measured for Oil-Red-O dissolved in methyl ethyl ketone (MEK) and an aqueous Acid-Red-1 solution. (a) Linear absorption coefficient~$\alpha$ as function of the mass fraction $w$ of dissolved dye. The transmission measurements of an aqueous Acid-Red-1 solution (circle marker~\refl{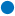}) confirm $\alpha = \alpha_0\, w$ to be valid up to the solubility limit $w_s = 0.05$ as reported by the manufacturer, i.e.\ for $w \le w_s$. At even higher concentration the absorption coefficients saturates for $w \ge w^*_s = 0.09$, which was confirmed with a Z-scan technique (circle marker~\refl{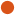}) by \citet{gayathri_single-beam_2007}. The solid line (\refl{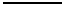}) is the linear ($w \le w^*_s$) and constant ($w > w^*_s$) fit, where our impact experiments are performed in the unsaturated regime. Preliminary transmission measurements in a limited range of $w$ for Oil-Red-O dissolved in MEK (square marker~\refl{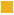}) yield similar results (see also \Tab\ref{tab:FluidProperties}). \label{fig:Absorption} (b) Surface tension as function of time $t$ during pendant drop measurements. The time $t\approx 0\,\mathrm{s}$ corresponds the moment the drop is formed at the tip of a capillary tube. The surface tension of each solution $\gamma$ is compared to the surface tension of the pure solvent $\gamma_0$ for three different solutions: an aqueous Acid-Red-1 solution at low ($w=0.002\,w_s$, circle marker~\refl{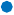}) and high ($w=0.75\,w_s$, circle marker~\refl{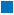}) mass concentration of dye, as well as Oil-Red-O dissolved in MEK at approximately half the solubility limit ($w=0.5\,w_s$, square marker~\refl{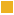}).\label{fig:SurfaceTension}} 
\end{figure}

We measured the absorption coefficient for the aqueous Acid-Red-1 solution over a large range of mass fractions of dye $w$ and find $\alpha = \alpha_0\, w$ with $\alpha_0 = (8.05\pm0.2)\times 10^{6}\, \mathrm{m}^{-1}$, compare \Fig\ref{fig:Absorption}\,(a). Our results are based on measuring the transmitted energy through liquid samples of known optical path length and are in very good agreement with measurements performed at the solubility limit with a Z-scan technique~\cite{gayathri_single-beam_2007}. To conclude, the aqueous Acid-Red-1 system allows to vary the laser-matter interaction in the range $5\times 10^{-5} \le \Ro\,\alpha \le 400$, where the lower limit is given by pure water and the upper limit by the solubility limit $w_s = 0.05$ as stated by the manufacturer.

A second solution consists of methyl ethyl ketone (MEK, CASRN:\ 78-93-3, product number 04380 by Sigma-Aldrich) as the solvent and Oil-Red-O (CASRN:\ 1320-06-5, product number 1320-06-5 by Sigma-Aldrich) as the dye. The solvent MEK exhibits a lower latent heat of vaporization as compared to water, which is expected to cause an even stronger drop response in terms of induced flow velocities for a given normalized penetration depth, i.e.\ ${(\Ro\,\alpha)}^{-1}$, and laser energy~\cite{klein_drop_2015}.\@ Initial experiments indeed show that this solution also operates at $\Ro\,\alpha \gg 1$ but with a much stronger response as compared to any aqueous Acid-Red-1 solution. Consequently, the same liquid motion can be induced at lower laser-pulse energy for dyed MEK as compared to an aqueous solution. Improvements of the laser-beam profile, which lead to a decrease in available laser-pulse energy at the drop impact location, can then be applied without limitations in the fluid-dynamic parameter space.

A liquid property that is very important in our experiments is the surface tension~$\gamma$, as the late-time fluid dynamics are determined by surface-tension forces in comparison to the kinetic energy that the drop gains upon laser impact~\cite{klein_drop_2015}.
Even the smallest amounts of surfactants may affect the surface tension of a liquid sample. Since we cannot exclude surface-active impurities in our solutions, i.e.\ due to left-overs from the dye manufacturing process, we measure the surface tension for both liquid systems. Pendant-drop measurements~\cite{de_gennes_capillarity_2002} are performed in a contour-analysis system (OCA by DataPhysics Instruments) that allows to record the surface tension of the solution in the pendant drop as function of time, see \Fig\ref{fig:SurfaceTension}\,(b). The results show that on timescales relevant to the laser impact on a drop, where the largest timescale is of the order of a few milliseconds, the surface tension of the solution can be approximated by the value of the pure solvent within $\pm 2\,\%$, see \Tab\ref{tab:FluidProperties} for the measured values.
However, the surface tension decreases by more than $30\,\%$ for the aqueous Acid-Red-1 solution at high concentration during pendant-drop measurements that last $20\,\mathrm{min}$ or even longer (not shown in the figure).
Such long measurement durations are excluded for the volatile MEK solution, which explains why the measurement is stopped after $30\,\mathrm{s}$ in \Fig\ref{fig:SurfaceTension}\,(b).

The solvent MEK requires the drop-impact chamber to have an inert atmosphere, since it is a hazardous substance that can be ignited under ambient temperature conditions~\cite{sigma-aldrich_2-butanone_2016}.\@ A further complication is the rapid evaporation of MEK, which is partially due to its low latent heat of vaporization and which can affect the drop-generation control mechanism that was described in \Chap\ref{sec:Chamber}.\@
The evaporation of the drop at the tip of the capillary leads to an increase of dye concentration and, once the solubility limit is reached locally, causes an agglomeration of dye particles, an effect also known for the self-assembly of colloidal dispersions during spray drying~\cite{sen_evaporation_2009}.\@
These particles form structures around the tip of the capillary tube leading to an unsteady pinch-off behavior of the drop or, even worse, to a complete clogging of the capillary once the flow rate is too low.

We avoid the clogging and stain formation by an active feedback loop controlling the volume flow rate $\dot{V}$ based on the drop pinch-off frequency $\fexp$.\@ 
The syringe pump, which sets~$\dot{V}$, is not switched off between experiments but set to a low volume flow that ensures a drop pinch-off approximately every $20\,\mathrm{s}$.
Before an actual experiment, the feedback loop is started to tune the flow rate to match the design repetition rate $\fexp=1\,\mathrm{Hz}$ within $\pm 1\%$, which takes typically about one minute depending on the stain formation before the feedback loop has been started. This ensures that the capillary and its tip are flushed and residual dye particles are removed.
\begin{table}[h]
	\centering
	\caption{\label{tab:FluidProperties}Liquid properties of an aqueous Acid-Red-1 solution and Oil-Red-O dissolved in methyl ethyl ketone (MEK). The surface tension~$\gamma$ of the solution is unaffected by the addition of the dye on short timescale, i.e.\ $t< 1\,\mathrm{s}$, and can be approximated by the value of the pure solvent. Likewise, the density is given by the value of the pure solvent, which is valid for a low dye mass fraction, i.e.\ $w \ll 1$.}
	\begin{ruledtabular}
	\begin{tabular}{lcc}
		\emph{parameter} & H\textsubscript{2}O \& & MEK \& \\
	 & Acid-Red-1 & Oil-Red-O \\\midrule
		 	liquid density $\rho$ (kg/m$^3$) & 998 & 805 \\
		 	surface tension $\gamma$ (N/m) & 0.072 & 0.025\\
		  linear absorption &  & \\
		    coefficient $\alpha_0$ ($1/\upmu\mathrm{m}$)  & $8.05\pm0.2$ & $8.13\pm0.1$\\
	\end{tabular}
	\end{ruledtabular}
\end{table}

\section{System control and timing}\label{sec:Control}
The control system for the lasers, cameras, energy meters, and auxiliary devices such as the syringe pump or motorized stages needs to ensure a precise timing and deterministic control of the devices. However, the timescales that need to be respected by a certain device may differ as illustrated in \Fig\ref{fig:Devices}\,(a). The devices in the \emph{asynchronous group} do not communicate directly with each other but only via the control software running on a computer (Z420 workstation by Hewlett Packard, Windows 7 by Microsoft as operating system, control software is implemented in MATLAB by MathWorks). The precision of the timing between devices in this group is set by the combined processing time of the control software and operating system. Since we do not use any real-time operating system we cannot guarantee deterministic synchronization between the devices and, therefore, settings are applied in an asynchronous fashion.

By contrast, the \emph{synchronous control} group consists of hardware that needs to interact with each other deterministically on a timescale as small as a few nanoseconds. This synchronization is achieved by hard-wired triggers transmitted directly among devices by a transistor-transistor logic (TTL). The precise timing among signals and their logical relation is controlled by a programmable pulse-delay generator (BNC575 by Berkeley Nucleonics Corporation with extended firmware and a precision of $250\,\mathrm{ps}$). The pulse-delay generator is programmed before an actual set of experiments in an asynchronous fashion from the control software but it operates independently during the actual experiments. 

For example, during a stroboscopic measurement we may change the flow rate of the syringe pump but cannot know for sure when the setting is applied by the device due to residual flow in the long feed lines. In principle, this is not a problem, since the effect of the flow rate, in our experiment the repetition rate of the drop pinch-off, is measured by a frequency counter in a synchronized way. Similarly, when changing the position of a $\lambda/2$-plate in a motorized mount we can monitor its effect by an energy meter. However, to ensure correct synchronization between devices of the different groups, the experiment needs to be stopped first. Then, the settings need to be applied, while the state of any asynchronous device is read out until the new settings are confirmed. Only then, the experiment can be restarted for a known and synchronized setting.

The control and timing of the devices is described along \Fig\ref{fig:Devices} for a stroboscopic experiment as it poses the highest demands in terms of reproducibility and a precise synchronization:
\begin{enumerate}
	\item The function generator (33600A by Keysight) is synchronized to its internal clock feeding pulses at a fixed frequency of $1\,\mathrm{Hz}$ to the main laser and pulse-delay generator, both connected to channel~\texttt{A} (please note that \Fig\ref{fig:Devices} shows the case where the function generator is already synchronized to the drop generation). The actual output of the pulse-delay generator is off to prohibit any actual laser output or devices being activated. However, from this point onward the cavity of the main laser is pumped by its flash-lamp allowing for the temperature-stabilized cavity to reach thermal equilibrium.
	\begin{figure*}[t!]
		\begin{minipage}[t]{0.50\textwidth}
			\mbox{}\\
			\includegraphics[width=12.25cm]{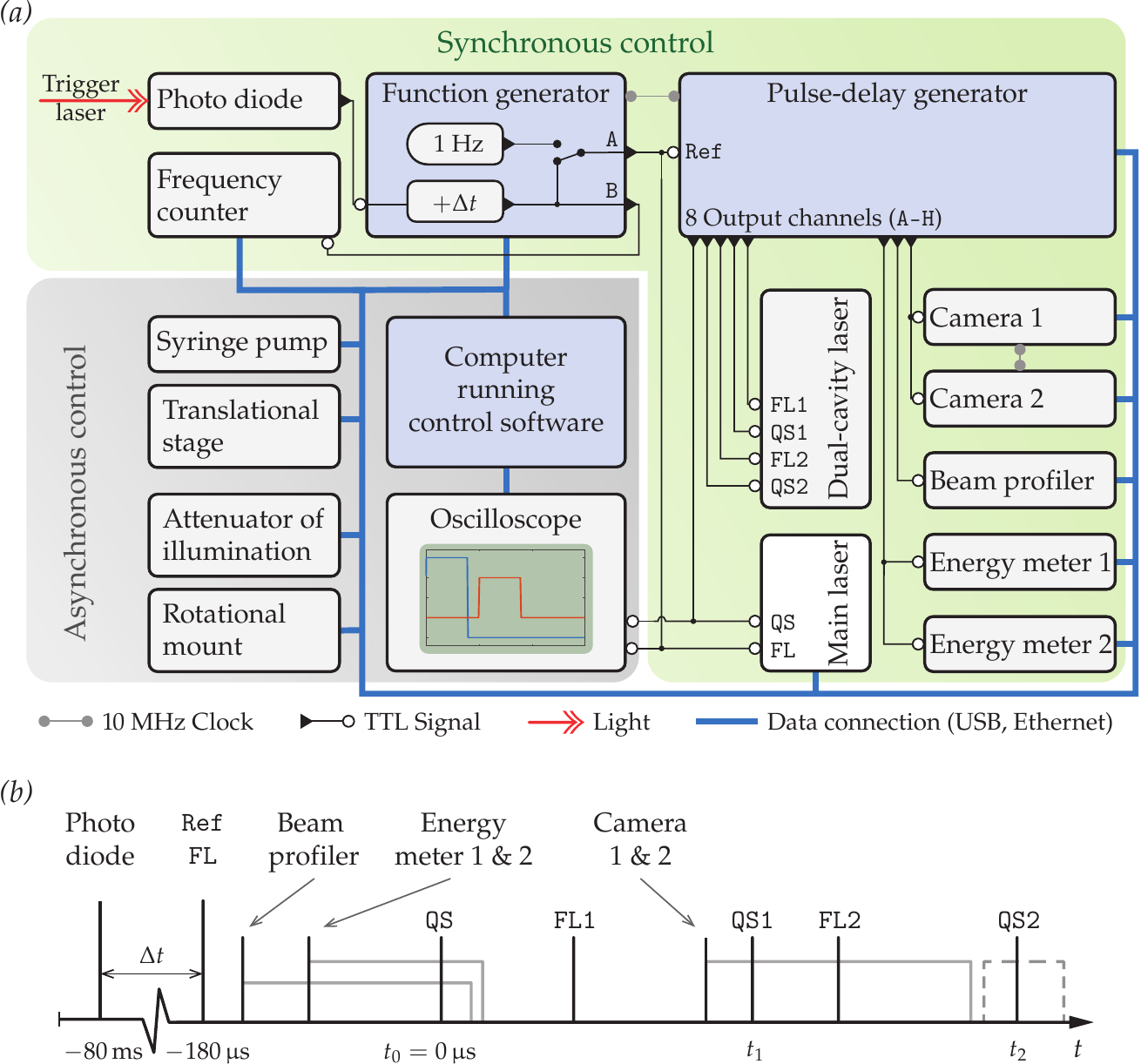}
			\end{minipage}\hfill
			\begin{minipage}[t]{0.2975\textwidth}
				\mbox{}\\[-\baselineskip]
				\caption{(a) Block diagram of the devices typically used in a laser-impact experiment. The diagram is split into two groups: the \emph{asynchronous group} holds devices that are controlled on a timescale much longer than a single experiment and do not require a real-time synchronization. On the other hand, the \emph{synchronous control} group consists of hardware that needs to interact with each other deterministically on a timescale down to a few nanoseconds. Our software to control devices in an object-oriented way in MATLAB and handle the data from the cameras or beam profiler is available online~\cite{klein_matlab-devices:_2017, klein_matlab-beamprofiles:_2017, klein_matlab-videos:_2017}. (b) Timing diagram of a stroboscopic experiment. Each vertical line represents the rising edge of a trigger signal. In case the pulse width of a signal is of importance, either because a device needs a certain activation time (beam profiler and energy meters) or reacts also to the falling edge of the signal (camera 1 and 2), the complete pulse is shown as gray solid line. The dashed line indicates the second exposure of the dual-frame camera, which is a response to the falling edge of its TTL signal.\label{fig:Devices}} 
			\end{minipage}
		\end{figure*}
	\item The syringe pump is switched on, drop generation starts and each drop generates a trigger signal at the photodiode that is received by the function generator. The signal is fed back to a frequency counter (J-50MB-YAG-USB by Coherent, an energy meter used for testing or calibration but also capable of frequency counting) connected to channel~\texttt{B} of the function generator. 
	\item At this point the drop generation does not yet affect any other device (main laser and pulse-delay generator are still synchronized to a fixed frequency), allowing to optimize the syringe pump. An active feedback loop controls the flow rate of the pump until the design repetition rate of the experiment $\fexp = 1\,\mathrm{Hz}$ is reached within $\pm 1\%$, see \Chap\ref{sec:Liquids}.
	\item The pulse-delay generator is programmed to create appropriate trigger signals once it will be activated. Likewise, the energy meters and cameras are prepared by first flushing the onboard memory and followed by the initialization of a new data acquisition. Any motorized stage, such as the rotational mount of the $\lambda/2$-plate to set \ELaser, is moved into position.
	\item The output of channel \texttt{A} of the function generator is synchronized to the trigger received from the photodiode (\Fig\ref{fig:Devices} shows this case). At this point the flash-lamp of the main laser and the actual experiment itself are synchronized to the drop generation. The setup is initialized and the generation of timing signals can then be started at any time.
	\item The output of the pulse-delay generator is activated leading to an output of timing signals as illustrated in \Fig\ref{fig:Devices}\,(b). Each time a drop masks the photodiode a trigger signal is generated at $t=-80\,\mathrm{ms}$. A delay $\Delta t$ is added to that trigger by the function generator before it is passed to the pulse-delay generator as reference trigger \texttt{Ref} and to the flash lamp \texttt{FL} of the main laser. The flash-lamp signal \texttt{FL} needs to precede the actual output (activated by the Q-switch signal \texttt{QS}) of the main laser by about $180\,\upmu\mathrm{s}$. The moment the main laser receives \texttt{QS} defines the time of laser impact $t=0$, at which point the drop needs to be at the desired impact position. The alignment along $\vec{e}_x$ can be tuned by setting an appropriate $\Delta t$ (alignment along $\vec{e}_y$ and $\vec{e}_z$ is done mechanically, see \Chap\ref{sec:Chamber}). The beam profiler and energy meters require a certain activation time, which is why they receive a trigger signal before \texttt{QS}. The time an image is taken is set by the moment of exposure by the pulsed light source. In case the dual-cavity laser as described in \Chap\ref{sec:Visualization} is used, the exposure of the first and second image is set by the Q-switch signal \texttt{QS1} and \texttt{QS2}, respectively. Also here, the laser pumping by the corresponding flash-lamp needs to be activated before the Q-switch signal is received (about $135\,\upmu\mathrm{s}$ for our laser system). It is important to note that the earliest possible time to generate an output for the pulse-delay generator is immediately after receiving trigger \texttt{Ref}, which leaves $180\,\upmu\mathrm{s}$ before the laser impact to activate all other devices. Given the laser-specific timings this restriction allows us to take stroboscopic images starting at $t=-45\,\upmu\mathrm{s}$. 
	\item The experiment of the single-drop impact is repeated for as many drops as requested before the output of the pulse-delay generator is stopped. The back-view camera can be moved along $\vec{e}_z$ during the experiments to tune the image plane to the position of the drop in each frame. Where possible, the data acquired by the devices is already streamed during the experiment to a solid-state disk, but the control software checks after the trigger signals are stopped whether an equal number of records are acquired from all synchronized devices. The oscilloscope can be used during the experiment as live visualization to check the timing signals. 
\end{enumerate}
	
The triggering changes slightly when the high-speed cameras are used as they record for each experiment all events from the laser impact at $t=0\,\mathrm{s}$ to the late-time events at a fixed frame rate, typically $f_\mathrm{cam}=10^4\,\mathrm{fps}$ at maximum image resolution.
Therefore, the camera trigger is synchronized to the \texttt{QS} trigger (\Fig\ref{fig:Devices}\,(b)). However, this synchronization does not yet ensure that the first frame is actually exposed at $t=0\,\mathrm{s}$ as the internal clock of each high-speed camera is freely running leading to a maximum uncertainty in the frame time of $1/f_\mathrm{cam}=100\,\upmu\mathrm{s}$.
In order to ensure proper synchronization to the laser impact we therefore reset the internal clock of each high-speed camera just before the laser pulse is emitted at $t=0\,\mathrm{s}$, also called a random-reset mode. This procedure then allows to take high-speed images with a precision of a few nanoseconds relative to the laser impact.

This section ends the description of the experimental apparatus. Our software to control devices in an object-oriented way in MATLAB and handle the data from the cameras and beam profiler is available online as open source~\cite{klein_matlab-devices:_2017,klein_matlab-beamprofiles:_2017,klein_matlab-videos:_2017}. We then used the setup to obtain the results presented in the next section: the stability of the control parameters of our experiment is discussed, followed by examples of the stroboscopic and high-speed imaging.

\section{Results}\label{sec:Results} 
The purpose of the apparatus is the controlled and quantitative study of the fluid-dynamic response of the drop to the laser impact. The apparatus has been used at different stages of construction to acquire the data presented in previous publications for water and solvent drops~\cite{klein_drop_2015,klein_laser_2015,gelderblom_drop_2016,kurilovich_plasma_2016}.\@ Examples for all visualization techniques described in this paper can be seen in two award-winning videos for the side-view visualization~\cite{klein_milton_2014-1, klein_fine_2016-1}.
Our studies focus on the drop propulsion, the change of liquid topology from the initial sphere to a thin sheet, and the simultaneous breakup of the deforming liquid body into smaller fragments. The study of the fragmentation poses the highest demands on the experimental setup. The need for a high spatial and temporal resolution of the visualization makes stroboscopic imaging an essential method in our studies, which in turn requires the experiment to be performed successively under constant control parameters. However, the fragmentation in liquids is often caused by instabilities~\cite{villermaux_fragmentation_2007} that amplify initial perturbations, which can be as small as the thermal noise in the system~\cite{eggers_physics_2008}. Therefore, the final fragmentation of the drop upon laser impact is inevitably subject to statistical fluctuations.
High-speed imaging is then necessary to visualize both the fragmentation and the entire evolution of the drop for a single realization of the experiment. 

In this section, we give a few examples based on the laser impact on a drop that illustrate the capabilities of the apparatus. We first show the stability of the control parameters of the apparatus in \Chap\ref{sec:Stability} followed by a brief introduction to the laser impact on a drop based on the stroboscopic imaging in \Chap\ref{sec:ImpactRegimes}.
We introduce the high-speed imaging in \Chap\ref{sec:RotateTheKick} for experiments that illustrate how the fluid dynamics can be controlled by the laser-beam profile. 

\subsection{Laser and drop stability}\label{sec:Stability}
The independent control parameters of our experiment are the spatial distribution of laser energy per unit area expressed as fluence $F(x,y,z)$, the liquid of the drop, and the initial radius \Ro\ and position of the drop in the lab reference frame. The liquid refers here to a specific solution that sets the absorption coefficient $\alpha$ and liquid properties such as density~$\rho$ and surface tension~$\gamma$ as already explained in \Chap\ref{sec:Liquids}. These quantities characterize a specific laser-impact experiment and need to be known for each realization. A dependent quantity or outcome of the experiment such as the propulsion speed of the drop due to the laser impact can be inferred in the post-processing from the recorded data. The information combined finally leads to the relevant physical dimensionless parameters, such as the Weber number, which characterizes the fluid-dynamic response in a general way, independent of the actual experimental realization. In this section, we focus on how we determine the independent control parameters and evaluate their stability over time, which is of great importance for the stroboscopic imaging.

The beam focusing and imaging described in \Chap\ref{sec:Description} are performed on a length scale in $\vec{e}_z$ that is large compared to the drop radius, e.g.\ $\f{5}/\Ro \gg 1$. Therefore, the fluence $F(x,y,z)$ can assumed to be constant as it propagates along $\vec{e}_z$ across the drop-impact location $-\Ro \le z \le \Ro$. The spatial distribution is imaged in relative terms $f(x,y,z=0)$ by the beam profiler as described in \Chap\ref{sec:Laser}. The total energy of the laser beam \ELaser\ at the drop-impact location is inferred from an appropriate calibration of energy meter 1. Both measurements combined allow to calculate the fluence in absolute terms 
\begin{eqnarray}
	F{(x,y,z=0)} &=& F_0\,f{(x,y,z=0)},\ \text{with}\\
	F_0 &=& E\,{\left(\int_A f{(x,y,z=0)}\ \mathrm{d}A\right)}^{-1},\label{eq:IntegrateBeam}
\end{eqnarray}
where $A$ corresponds to an area on the CCD of the beam profiler large enough to cover the complete beam. Care must be taken to calibrate the base level of the beam profiler to exclude the noise in the integration of \Eq(\ref{eq:IntegrateBeam}). To determine the size of the laser beam and the corresponding area $A$ we follow the ISO standard~\cite{_iso_2005-1}; our implementation is available online~\cite{klein_matlab-beamprofiles:_2017,klein_matlab-videos:_2017}.

\begin{figure}
	\centering
	\includegraphics{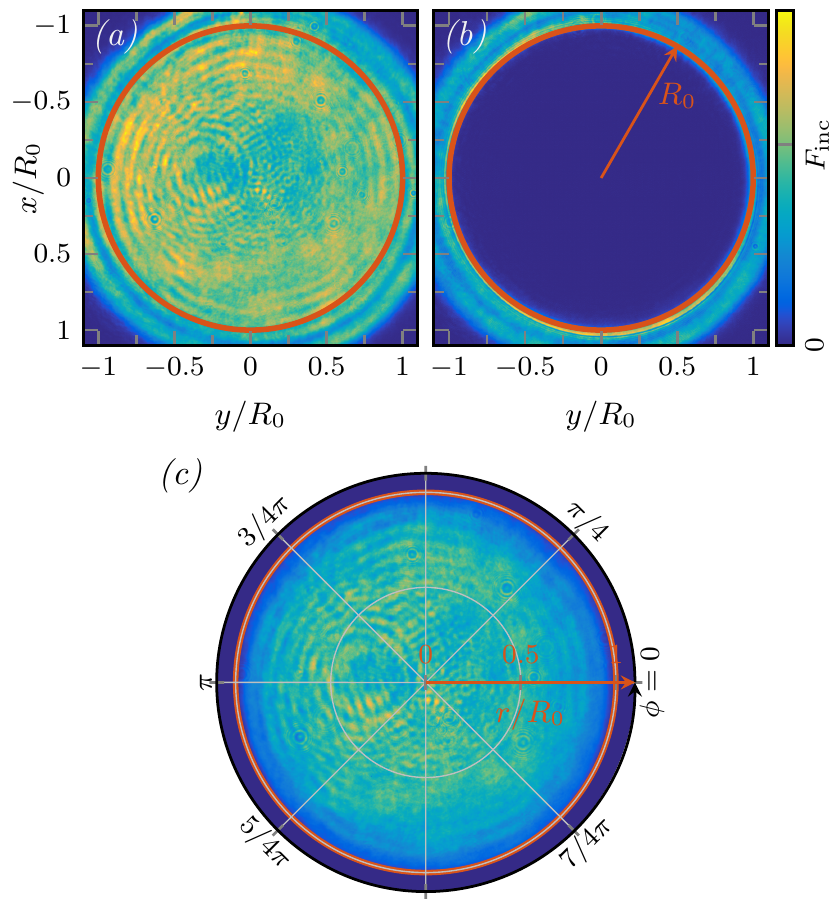}
	\caption{Laser-beam profile recorded in absence of the drop in (a) and with the drop in (b). The solid line (\refl{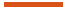}) in both images indicates the drop size and position determined from (b). This information allows to calculate the energy per unit area $F_\mathrm{trans}$ transmitted into the drop as shown in (c) taking Fresnel reflection~\cite{hecht_optics_2002} at the liquid-air interface into account. The quantity $F_\mathrm{inc}$ shown in the colorbar is the mean fluence in the part of the laser beam that the drop is exposed to. The diffraction pattern in the measured beam profile is related to the resonator modes of the laser cavity and remaining liquid fragments on lens \f{6} (see \Fig\ref{fig:Chamber}) from preceding experiments, leading to a standard deviation of the fluence that is incident on the drop of $16\,\%$.\label{fig:Beamprofile}} 
\end{figure}

\begin{figure}[t]
	\centering
	\includegraphics{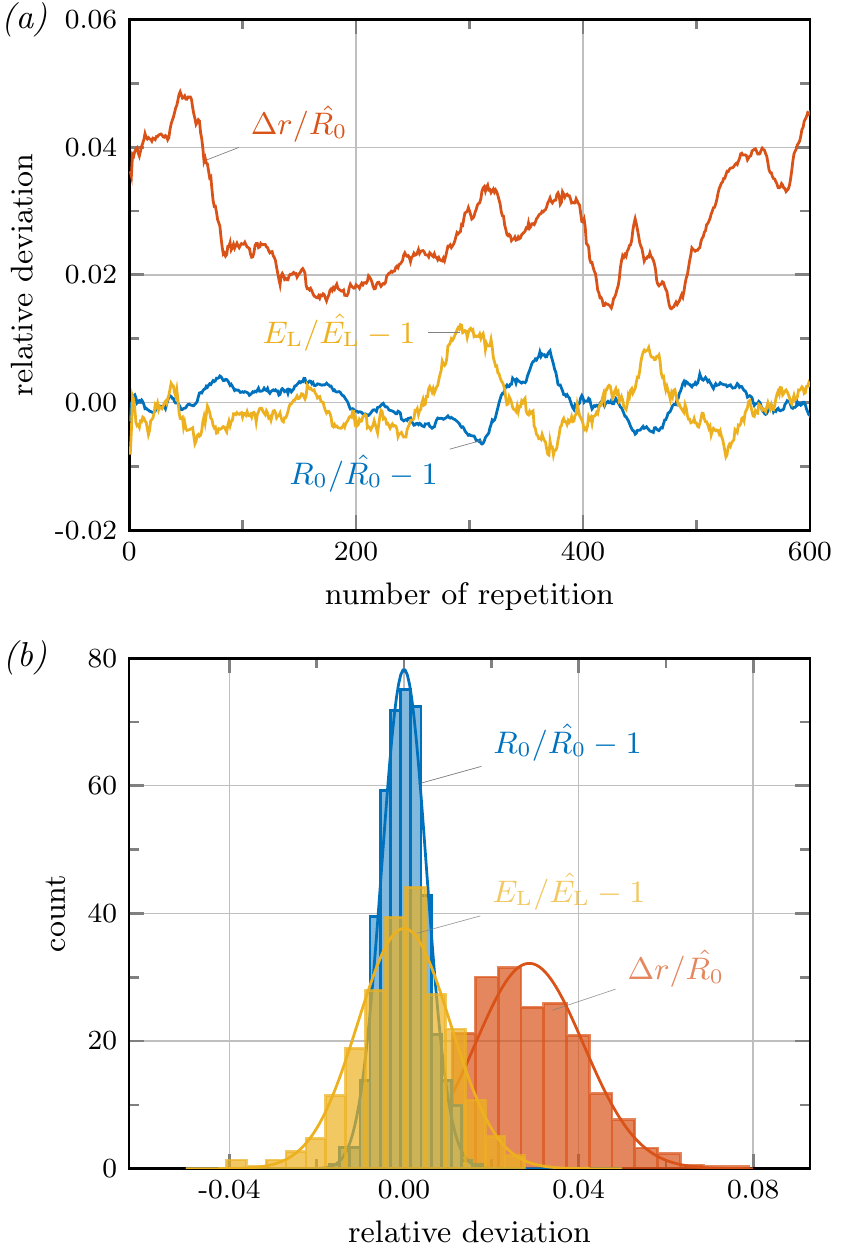}
	\caption{The relative deviation among 600 repetitions over the course of 10 min in (a) and as histogram in (b) for the laser energy \ELaser, initial drop radius \Ro, and the offset $\Delta r$ in the radial position of the drop. Each quantity is normalized by an appropriate mean value, either the mean initial radius $\hat{R}_0$ or the average laser energy $\hat{E}_\mathrm{L}$. The solid lines in (b) represent a Gaussian fit to each distribution yielding the standard deviation $\sigma$ for each control parameter: $\sigma_E = 0.01$, $\sigma_{\Ro} = 0.005$ with an average relative offset in the drop position of $ 0.0287 \pm 0.0124$.
	 The data is taken from the experiment shown in \Fig\ref{fig:Stroboscopic} and a running average filter is applied to the data in (a) for better visualization (statistics in (b) are based on the unfiltered data).\label{fig:Stability}}
\end{figure}

A typical beam-profile is shown in \Fig\ref{fig:Beamprofile}\,(a) obtained for the optical path from \Chap\ref{sec:Near-fieldImaging} including the near-field imaging and beam-shaping technique. The measurement is performed at low energy to avoid plasma formation in the intermediate focus at $z=30\,\mathrm{mm}$ (see \Fig\ref{fig:Chamber}). 
During each realization of the experiment, the image recorded by the beam profiler can be used to determine the drop position and radius \Ro\ as shown in \Fig\ref{fig:Beamprofile}\,(b). We obtain good results even though a plasma may be created in the intermediate focus behind the drop: the optical breakdown occurs close to the intermediate focus at $z>30\,\Ro$ and is not within the depth of field of the imaging optics of the beam profiler. We find a spread in \Ro-values between measurements based on the beam-profiler image data and camera recordings of the side-view and back-view to be within $\pm 3\%$. Combining the information of \Ro, the drop position, and the fluence we can compute the energy transmitted into the drop per unit area (\Fig\ref{fig:Beamprofile}\,(c)). Moreover, the use of the laser beam to not only induce the liquid motion but also to image the initial drop is advantageous during stroboscopic recordings: the side- and back-view imaging can then be used record the late-time response of the drop for a known initial radius~\Ro.

The beam profile as given in relative terms $f(x,y,z=0)$ is not a function of the total laser energy, which we confirmed for the relevant range $E_\mathrm{L,min} \le \ELaser \le E_\mathrm{L,max}$. It is therefore sufficient to measure the beam profile once for a given focusing or imaging condition and only measure the scalar quantity \ELaser\ for each experimental realization to compute the absolute scale according to \Eq(\ref{eq:IntegrateBeam}). In case the near-field imaging is not used but the multiple modes of the beam propagate out-of-phase to the drop-impact location, the spatial distribution is deteriorated as can be seen by comparison of \Fig\ref{fig:Beamprofile}\,(a) and \Fig\ref{fig:RotateTheKick}\,(a,~b). As higher modes of the laser beam decay faster the superposition of all modes changes along $\vec{e}_z$. This effect illustrates the advantage of an imaging technique where the modes in the conjugated focal plane preserve their phase relation.

The stability of the control parameters over time is shown in \Fig\ref{fig:Stability} for the same stroboscopic experiment visualized in \Fig\ref{fig:Stroboscopic}.\@ The liquid in this experiment is a MEK solution at about half the solubility limit and the laser energy $\ELaser=113\,\mathrm{mJ}$, which leads to a violent breakup of the drop into tiny fragments. This operation point represents a demanding test case for the apparatus as many fragments are created during each realization that could possibly interfere with subsequent repetitions of the experiment. The control parameters of the experiment are stable over time as quantified by the standard deviation $\sigma$ obtained from a Gaussian fit to the distribution of each control parameter, see \Fig\ref{fig:Stability}.\@ To conclude, the setup allows to measure the control parameter for each experimental realization accurately and the shot-to-shot variation of the control parameters is less than $3\%$.

\subsection{Laser impact regimes}\label{sec:ImpactRegimes}
\begin{figure*}
	\centering
	\includegraphics{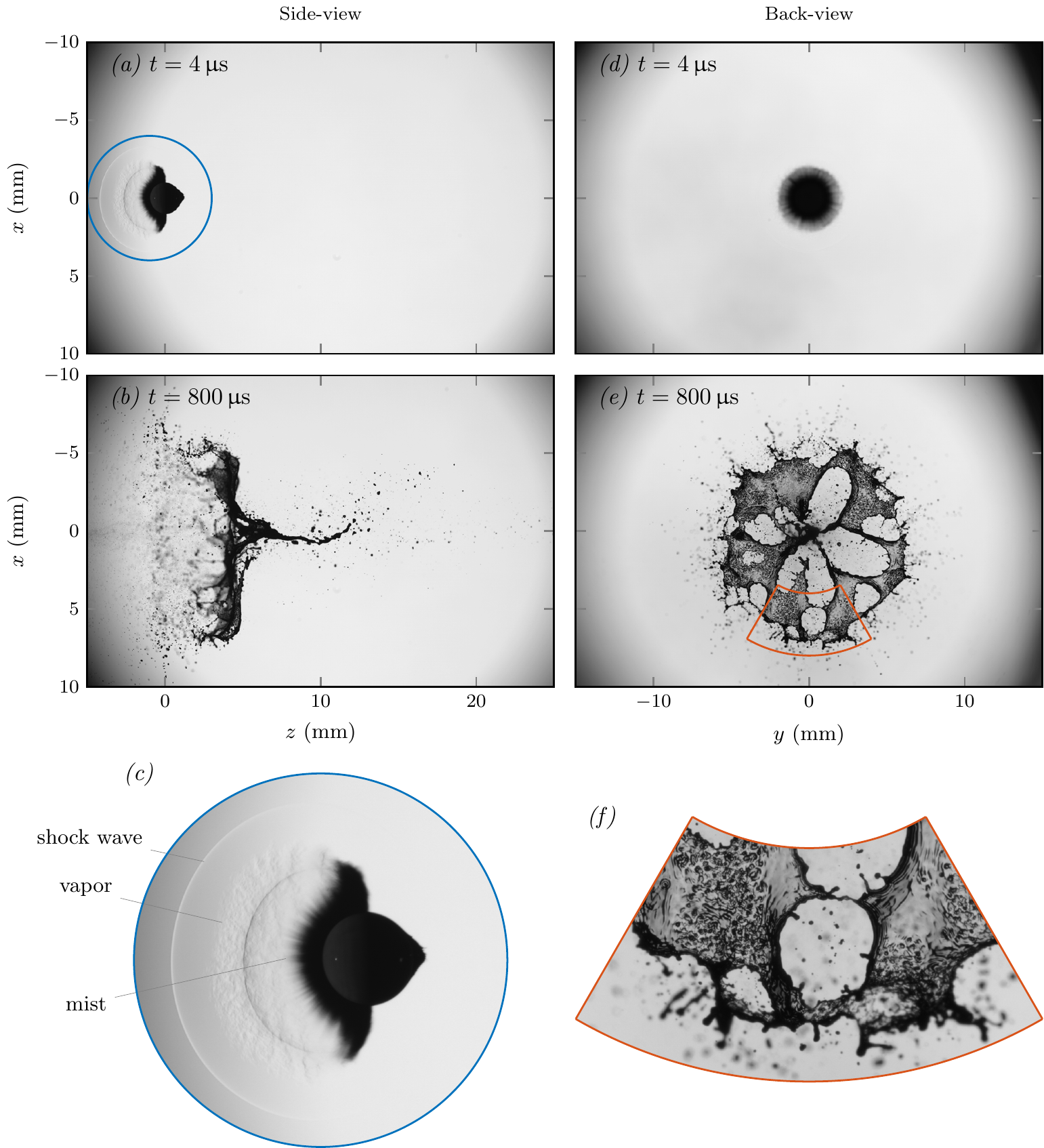}
	\caption{Side-view (a--c) and back-view (d--f) images of the drop taken stroboscopically at $t=4\,\upmu\mathrm{s}$ (a, c, d) and $t=800\,\upmu\mathrm{s}$ (b, e, f) after laser impact. Images (c) and (f) are magnified regions of the full-frame images (a) and (e). The visualization is performed with the monochromatic and pulsed light source shown in \Fig\ref{fig:Optics}. The laser energy per pulse is $\ELaser=113\,\mathrm{mJ}$ and the liquid is Oil-Red-O dissolved in MEK at approximately half the solubility limit ($w=0.5\,w_s$). The images show the fluid-dynamic response of the drop to the laser impact at two characteristic points in time. The early-time recordings (a, d, c) show a spherical shockwave emitted at the drop surface, followed by the ejection of mass in liquid and vapor state. A second shockwave travelling towards the drop surface is visible inside the vapor phase as dark region (see text). The late-time consequence of the laser impact (b, e, f) is a change of the liquid topology into a thin sheet (visible as the liquid is semi-transparent in the visualization) that breaks to form tiny and stable drops.\label{fig:Stroboscopic}} 
\end{figure*}

The impact of a laser pulse on a liquid drop can lead to a violent response as already shown in \Fig\ref{fig:ImpactRegimes}: the liquid is set into motion on a timescale of a few microseconds, strongly deforms, and eventually fragments. The first reaction to the laser impact visible in the figure is the emission of a shock wave in the surrounding air (\Fig\ref{fig:ImpactRegimes}\,(a) and \Fig\ref{fig:Stroboscopic}\,(a, c)). The shock wave is produced by the ejection of matter in a vapor and liquid state: the shock wave, visible as bright half-circle at the very left of \Fig\ref{fig:Stroboscopic}\,(c), is followed by a region filled with vapor, which is visible as Schlieren in the images. Still attached to the left of the drop is a mist cloud, i.e. a two-phase mixture of vapor and tiny liquid drops, that is visible as a gray-to-black haze. A second shockwave is visible inside the vapor phase as dark region that is caused by the collision between the expanding vapor and the ambient air~\cite{vogel_sensitive_2006}.
Both shockwaves and the vapor plume are only visible at the periphery of the FOV as the local cut-off of the background light source leads to an oblique illumination~\cite{settles_schlieren_2001} that visualizes density gradients as intensity differences. As discussed in \Chap\ref{sec:Visualization}, the addition of a Schlieren technique would allow for a more detailed visualization of these phenomena. Eventually, the drop propels forward in the direction of the laser propagation $\vec{e}_z$ as the vapor is ejected in opposite direction $-\vec{e}_z$~\cite{klein_drop_2015}. The center-of-mass speed of the drop that is induced in the example of \Fig\ref{fig:Stroboscopic} is approximately $U=10\,\mathrm{m/s}$.\@ This speed can be further increased as the maximum laser energy is not fully used in this example.

The color images in \Fig\ref{fig:ImpactRegimes} are filtered by a notch filter centered at $\lLaser = 532\,\mathrm{nm}$, which leads to the magenta colored background in the images. Nevertheless, the color camera resolves the visible spectrum of the light and helps to conclude on the driving mechanism of the liquid motion, i.e.\ whether plasma formation and the corresponding plasma pressure needs to be taken into account. In case of a tightly focused beam (\Fig\ref{fig:ImpactRegimes}\,(a)), which is realized by moving lens \f{5} (\Fig\ref{fig:Optics}) such that the focal point of the lens is aligned with the drop, a white glow in the images reveals the formation of a plasma: the combination of different wavelengths in the visible spectrum appears white in the color images and indicates that the threshold for optical breakdown in the liquid is reached~\cite{kennedy_laser-induced_1997}. In this case, the plasma pressure then needs to be included in a description of the driving force for the liquid motion. When the beam waist of the laser is moved away from the drop by moving lens \f{5} closer to the drop, the fluence at the drop location is decreased below the threshold for plasma formation in water of about $100\text{ to } 400\,\mathrm{J/cm}^2$~\cite{vogel_mechanisms_2003}. Still captured by the camera is the fluorescence emission at a wavelength $\lambda > 532\,\mathrm{nm}$ (visible as yellow spot in \Fig\ref{fig:ImpactRegimes}\,(b,~c)), which is caused by an emission from a resonantly excited state of the dissolved dye molecules. The volumetric energy density in the superficial layer of the drop, which is a direct result of the local laser fluence and the absorption characteristics of the liquid-dye solution, is sufficient to induce local boiling~\cite{vogel_mechanisms_2003}. This phase change is then the only driving mechanism for the liquid motion in this case as the optical radiation pressure\footnote{The typical impulse $I$ exerted on the drop by the 3\% reflected light from the surface scales as $I\sim E_\mathrm{r}/c$, with $E_\mathrm{r}$ the energy of the reflected light and $c$ the speed of light. This impulse would yield a typical drop speed $U\sim 10^{-7}$ m/s. The impulse due to thermal radiation from the hot drop surface scales as $I\sim \epsilon \sigma T^4 \Ro^2 \taup/c$ with $\epsilon$ the emissivity and $\sigma$ the Stefan-Boltzmann constant. This impulse would yield $U\sim 10^{-14}$ m/s.} from the laser and the thermal radiation pressure caused by the heating of the drop surface are insignificant~\cite{delville_laser_2009}.

Care must be taken when interpreting the time an event occurs in stroboscopic images. All frames are exposed by a pulse of light less than $10\,\mathrm{ns}$ in duration at the time $t_1$ marked by \texttt{QS1} in the timing diagram~\ref{fig:Devices}\,(b).\@ This pulse of light captures the fluid dynamics free of motion blur on the CCD of the camera at a well-known point in time. However, the camera is activated sufficiently earlier in time to account for any delay of the electronics or even a mechanical shutter in front of the CCD.\@ Therefore, the CCD captures events also before and after $t_1$. In case of \Fig\ref{fig:ImpactRegimes} this includes any plasma glow or fluorescent-light emission during and shortly after the laser pulse. In fact, the camera is exposed twice: once by the plasma or fluorescence light emission at $t=0\,\mathrm{s}$ and later at $t_1$ by the pulsed light source.

The grid arrangement with a common axis for \Fig\ref{fig:Stroboscopic}\,(a, b, d, e) is possible due to a proper image calibration: each individual view is calibrated by its own set of calibration images and the spatial relation among the different views is inferred from the position of the initial drop. It is then crucial to record the drop before the laser impact to acquire an image of the drop free of any deformation. In that case, the spherical drop can be analyzed in the images of each view to find its center-of-mass position, which is then used as an unique reference point for the calibration.
A clear advantage of stroboscopic imaging is seen in \Fig\ref{fig:Stroboscopic}: events at arbitrary points in time can be captured even though they take place at times separated by several orders of magnitude. At early times, the side-view gives insight in the propulsion mechanism of a drop upon laser impact (see \Fig\ref{fig:Stroboscopic}\,(a, c)), whereas at late times details of the fluid dynamics such as corrugations on the evolving liquid sheet are revealed in \Fig\ref{fig:Stroboscopic}\,(e, f).\@
In these recordings, the image resolution of $4008 \times 2672$ pixels per image is sufficient to resolve even small details in an overall large FOV.\@ 

\subsection{Controlling the fluid dynamics}\label{sec:RotateTheKick} 
A key feature of the experimental setup is its capability to control and image the spatial distribution of the laser radiation that drives the fluid dynamics. However, the timescale of laser radiation and drop dynamics are separated by several orders of magnitude. The question then arises: how reproducible is the late-time drop dynamics and can it be influenced by the early-time laser pulse? High-speed imaging plays an important role in the answer to this question since it does not require a reproducible process as in the case of stroboscopic imaging. Each individual feature can be followed in time for each realization of an experiment, only limited by the frame rate and image resolution. The two high-speed cameras allow to record images at a maximum resolution of $1024 \times 1024$ pixels at a frame rate of $f_\mathrm{cam} =10\,000\,\mathrm{fps}$.

\begin{figure}
	\centering
	\includegraphics{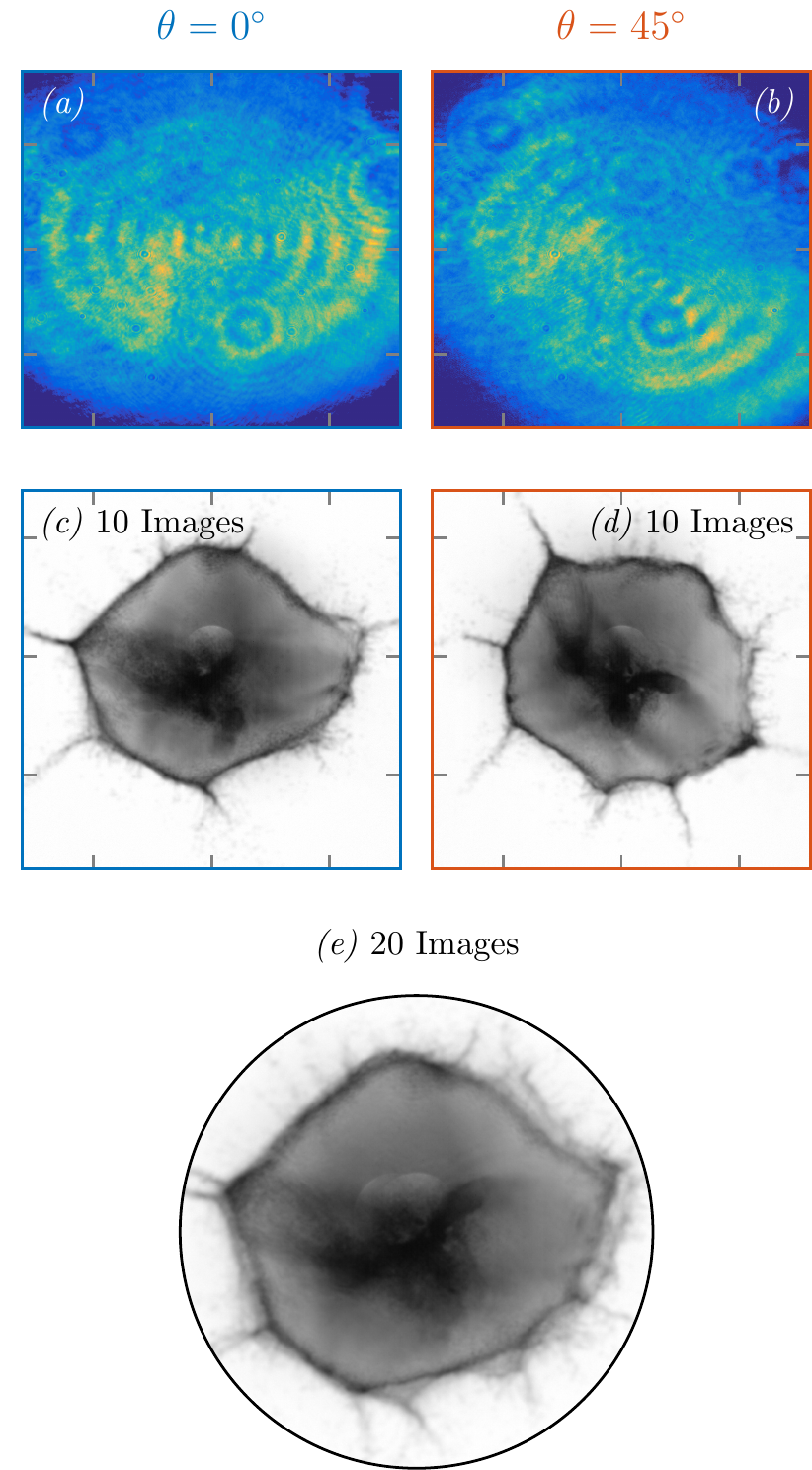}
	\caption{The influence of the laser-beam profile onto the fluid dynamics of the drop illustrated for two laser-beam profiles: a reference beam at a rotation about $\vec{e}_z$ of $\theta = 0^\circ$ in (a) and a rotated beam at $\theta = 45^\circ$ in (b). Ten experiments are recorded with a high-speed camera for each rotation $\theta$ and the superimposed frames taken at $t=900\,\upmu\mathrm{s}$ are shown in (c) and (d). The image (e) is the superposition of (c) and (d), where (d) as been rotated by $-45^\circ$ to compensate the initial rotation of the laser-beam profile. The liquid in this experiment is an aqueous Acid-Red-1 solution, resulting in a Weber number of $\We = \rho \Ro U^2 / \gamma = 120$ and a Reynolds number of $\Rey = \Ro U / \nu =2800$ with $\nu$ as the liquid kinematic viscosity and $U$ the propulsion speed of the drop.\label{fig:RotateTheKick}} 
\end{figure}

To answer the question on the reproducibility of the drop dynamics we record ten realizations of a laser-impact experiment at constant control parameters. The laser-beam profile is shown in \Fig\ref{fig:RotateTheKick}\,(a) and exhibits an elliptical shape, i.e.\ we do not improve the shape by the beam shaping described in \Chap\ref{sec:Near-fieldImaging}. The fluid-dynamic response to the laser impact can be observed in terms of the back-view visualization in \Fig\ref{fig:RotateTheKick}\,(c), which is a superposition of all ten realizations taken by the high speed recording at a fixed time. As can be seen, the shape evolution and even the position where ligaments are expelled radially outwards are reproducible features of the experiment: the gray value in the overlay image is a measure for the probability that ligaments occurred at the same position in all ten realization, where a black pixel means that in all experiments a ligament is found at a particular position. The advantage of high-speed imaging in this context is the ability to choose the point in time for the overlay in the post-processing after the experiment. Since the data of multiple recordings can be stored in the camera memory, all ten experiments can be recorded in direct succession within just ten seconds.

To investigate the influence of the laser-beam profile on the fluid-dynamic response, we now rotate the beam profile in \Fig\ref{fig:RotateTheKick}\,(a) by $\theta = 45^\circ$ to obtain the beam profile shown in \Fig\ref{fig:RotateTheKick}\,(b). Following the same procedure as before we come to the same conclusion: the laser impact on a drop shows a remarkable reproducibility. Moreover, a comparison of \Fig\ref{fig:RotateTheKick}\,(d) and (c) suggests that the drop dynamics follow the same rotation $\theta$ as set by the beam profile. This effect becomes clear when we overlay the two images, one corrected by the apparent rotation $\theta$, which is shown in \Fig\ref{fig:RotateTheKick}\,(e). 
The overlay reveals that the large-scale fluid dynamics is entirely controlled by the initial beam profile: the incident distribution of energy determines the volumetric energy density in the superficial layer of the absorbing drop after the laser impact, which induced a phase transition, expansion and recoil pressure, which in turn sets the driving force of the fluid dynamics. This result explains why the control and visualization of the beam profile is of great importance to our experiments: although the beam profile only acts on the drop during the very first $5\,\mathrm{ns}$, 
it determines the drop dynamics to a large extent at all later times and is therefore a crucial control parameter. 

\section{Summary}\label{sec:Summary}
	An experimental apparatus to control and visualize the fluid-dynamic response of a liquid target to a laser-induced phase change is presented. The laser is a Q-switched Nd:YAG laser system emitting pulses with a duration of $\taup=5\,\mathrm{ns}$ at a wavelength of $\lLaser=532\,\mathrm{nm}$ and a maximum energy of $E_\mathrm{L,max}=420\,\mathrm{mJ}$. We explain two optical arrangements to hit the liquid target of interest. First, an optical path is shown that focuses the freely-propagating beam onto the target. The advantages of this arrangement are its simplicity and options to extend the path easily. Second, an optical path to image the beam profile from the near-field of the laser onto the liquid target is explained in detail. The path is more difficult to implement but can be combined with a beam-shaping technique. The beam shaping that we present allows for any linear combination of the laser beam with a rotated version of itself, which is used to improve the axi-symmetry of the beam profile. 
	
A liquid drop with an initial radius $\Ro \approx 1\,\mathrm{mm}$ is generated as a target for the impact. We present two liquid-dye solutions that allow to tune the penetration depth of laser light into the drop in terms of a linear absorption coefficient~$\alpha$ over a wide range, i.e.\ $5\times 10^{-5} \le \Ro\,\alpha \le 400$. The response of the drop to the laser impact is visualized by high-speed and stroboscopic imaging in two orthogonal views: a side-view perpendicular to the laser beam and a back-view that is along the laser-beam propagation. The combination of imaging techniques with exposure times down to $10\,\mathrm{ns}$ resolves all relevant timescales that vary by orders of magnitude between the nanosecond duration of the laser impact and the millisecond evolution of the fluid-dynamic response.

The setup enables an operation at constant control parameters with a shot-to-shot variation of less than $3\%$, which makes stroboscopic imaging possible in the first place. We present examples from the laser impact on a drop that require this stability, foremost a case where the control over the beam profile allows perfect control over the late-time fluid dynamics.

\begin{acknowledgments}
We thank Chris Lee, Guillaume Lajoinie, Claas Willem Visser, Adam Lassise, Ri\"elle de Ruiter, Stefan Karpitschka, Henri Lhuissier, and Stefan B\"aumer for fruitful discussions. Furthermore, we acknowledge Koen Arens and Martijn van Gestel for their support in the absorption measurements.
This work is part of an Industrial Partnership Programme
of the Netherlands Organization for Scientific Research (NWO). This
research programme is co-financed by ASML.
\end{acknowledgments}

\bibliography{main} \end{document}